\documentclass[aps,prl,reprint,showpacs,superscriptaddress]{revtex4-1}
\usepackage[plainpages=false,pdfpagelabels,colorlinks=true,linkcolor=red,urlcolor=blue,citecolor=blue,pdftitle={Title},pdfauthor={},pdfdisplaydoctitle=true,pdfduplex=DuplexFlipLongEdge]{hyperref}
\usepackage{verbatim}
\usepackage[english]{babel}
\usepackage{bm}
\usepackage{amsmath}
\usepackage{epsfig}
\usepackage{array}
\usepackage{ulem}
\usepackage{color}
\usepackage{braket}
\usepackage{qcircuit}
\usepackage{xfrac}
\usepackage{listings}
\usepackage{framed}

\begin{document}

\title{ScQ cloud quantum computation for generating Greenberger-Horne-Zeilinger states of up to 10 qubits}

\author{Chi-Tong~Chen}
	\thanks{These authors contributed equally to this work.}
	\affiliation{Institute of Physics, Chinese Academy of Sciences, Beijing 100190, China}
    \affiliation{School of Physical Sciences, University of Chinese Academy of Sciences, Beijing 100190, China}

\author{Yun-Hao~Shi}
	\thanks{These authors contributed equally to this work.}
	\affiliation{Institute of Physics, Chinese Academy of Sciences, Beijing 100190, China}
    \affiliation{School of Physical Sciences, University of Chinese Academy of Sciences, Beijing 100190, China}

\author{Zhongcheng~Xiang}
	\thanks{These authors contributed equally to this work.}
	\affiliation{Institute of Physics, Chinese Academy of Sciences, Beijing 100190, China}
    \affiliation{School of Physical Sciences, University of Chinese Academy of Sciences, Beijing 100190, China}

\author{Zheng-An~Wang}
		\thanks{These authors contributed equally to this work.}
		\affiliation{Beijing Academy of Quantum Information Sciences, Beijing 100193, China}
	\affiliation{Institute of Physics, Chinese Academy of Sciences, Beijing 100190, China}

\author{Tian-Ming~Li}
	\affiliation{Institute of Physics, Chinese Academy of Sciences, Beijing 100190, China}
    \affiliation{School of Physical Sciences, University of Chinese Academy of Sciences, Beijing 100190, China}

\author{Hao-Yu~Sun}
	\affiliation{Institute of Physics, Chinese Academy of Sciences, Beijing 100190, China}
    \affiliation{School of Physical Sciences, University of Chinese Academy of Sciences, Beijing 100190, China}

\author{Tian-Shen~He}
    \affiliation{Department of Physics, Fudan University, Shanghai 200433, China}

\author{Xiaohui~Song}
	\affiliation{Institute of Physics, Chinese Academy of Sciences, Beijing 100190, China}
    \affiliation{School of Physical Sciences, University of Chinese Academy of Sciences, Beijing 100190, China}

\author{Shiping~Zhao}
\affiliation{Institute of Physics, Chinese Academy of Sciences, Beijing 100190, China}
\affiliation{School of Physical Sciences, University of Chinese Academy of Sciences, Beijing 100190, China}
\affiliation{CAS Center for Excellence in Topological Quantum Computation, University of Chinese Academy of Sciences, Beijing 100190, China}
\affiliation{Songshan Lake Materials Laboratory, Dongguan, Guangdong 523808, China}

\author{Dongning~Zheng}
  \email{dzheng@iphy.ac.cn}
\affiliation{Institute of Physics, Chinese Academy of Sciences, Beijing 100190, China}
\affiliation{School of Physical Sciences, University of Chinese Academy of Sciences, Beijing 100190, China}
\affiliation{CAS Center for Excellence in Topological Quantum Computation, University of Chinese Academy of Sciences, Beijing 100190, China}
\affiliation{Songshan Lake Materials Laboratory, Dongguan, Guangdong 523808, China}

\author{Kai Xu}
   \email{kaixu@iphy.ac.cn}
\affiliation{Institute of Physics, Chinese Academy of Sciences, Beijing 100190, China}
\affiliation{School of Physical Sciences, University of Chinese Academy of Sciences, Beijing 100190, China}
\affiliation{CAS Center for Excellence in Topological Quantum Computation, University of Chinese Academy of Sciences, Beijing 100190, China}
\affiliation{Songshan Lake Materials Laboratory, Dongguan, Guangdong 523808, China}

\author{Heng~Fan}
    \email{hfan@iphy.ac.cn}
\affiliation{Institute of Physics, Chinese Academy of Sciences, Beijing 100190, China}
\affiliation{School of Physical Sciences, University of Chinese Academy of Sciences, Beijing 100190, China}
\affiliation{CAS Center for Excellence in Topological Quantum Computation, University of Chinese Academy of Sciences, Beijing 100190, China}
\affiliation{Songshan Lake Materials Laboratory, Dongguan, Guangdong 523808, China}
\affiliation{Beijing Academy of Quantum Information Sciences, Beijing 100193, China}

\begin{abstract}
In this study, we introduce an online public quantum computation platform, named as ScQ, based on a 1D array of a 10-qubit superconducting processor. Single-qubit rotation gates can be performed on each qubit. Controlled-NOT gates between nearest-neighbor sites on the 1D array of 10 qubits are available. We show the online preparation and verification of Greenberger-Horne-Zeilinger states of up to 10 qubits through this platform for all possible blocks of qubits in the chain. The graphical user interface and quantum assembly language methods are presented to achieve the above tasks, which rely on a parameter scanning feature implemented on ScQ. The performance of this quantum computation platform, such as fidelities of logic gates and details of the superconducting device, are presented.
\end{abstract}

\maketitle

{\it Introduction.}---Quantum computers aim to realize quantum algorithms that may outperform classical computers. Recently, quantum supremacy and quantum advantage have been successfully demonstrated for the sampling output of random quantum circuits and Gaussian boson in laboratories~\cite{google,StrongQS,Photon}. Moreover, cloud (online) quantum computation (CQC) can be accessed worldwide and conveniently used for various aims, including research, application exploration and education. Some CQC platforms based on superconducting processors have been launched for use online, such as the IBM quantum experience (https://quantum-computin.ibm.com), on which a series of research have been performed~\cite{YuanhaoWang,IBMapp1,IBMapp2,IBMapp3,IBMapp4,Hollenberg,IBMGHZ,FNori,Yang2022}. It can be expected that CQC will be one of the main approaches for near term applications of quantum computation~\cite{Georgescu2014,Feynman,OneWayQComputer,QC1,HHL2017,YuanXiao2021,Hu2020,Hu2022,QML1,QML2,VQE1,QGAN,google1,Surfacecode,Shor1997,Gu2017,Xiang2013}. However, only limited resources of CQC are available.

As developers, we report a newly implemented superconducting CQC platform, dubbed as ScQ (ScQ cloud quantum computation platform is available at http://q.iphy.ac.cn, which is updated at Quafu cloud quantum computation platform \url{http://quafu.baqis.ac.cn}), currently equipped with a 1D array of 10 superconducting qubits. Single-qubit rotation gates can be performed on each qubit. Controlled-NOT (CNOT) gates on pairs of nearest-neighbor (NN) qubits on the chain are available. To demonstrate the performance of ScQ, we use a well-accepted benchmark, multi-qubit Greenberger-Horne-Zeilinger (GHZ) states generation of up to all 10 qubits, to show the capabilities and programming of our platform. Generally, the prepared GHZ states and other entangled states can be used as a valuable resource for various quantum computation tasks and testing principles of quantum theory~\cite{GHZ,RMP_Entanglement,RMP_MBEntanglement,PhysicsReport,JWPan18,MLHu,Friis18,GongM19,SongC17,SongC19,Monz,LukinScience,Wei2006,Matsuo2007}. These results together with gate information and device parameters, will be useful for researchers on ScQ. 

\begin{figure*}[ht]
\
\includegraphics[width=1\textwidth]{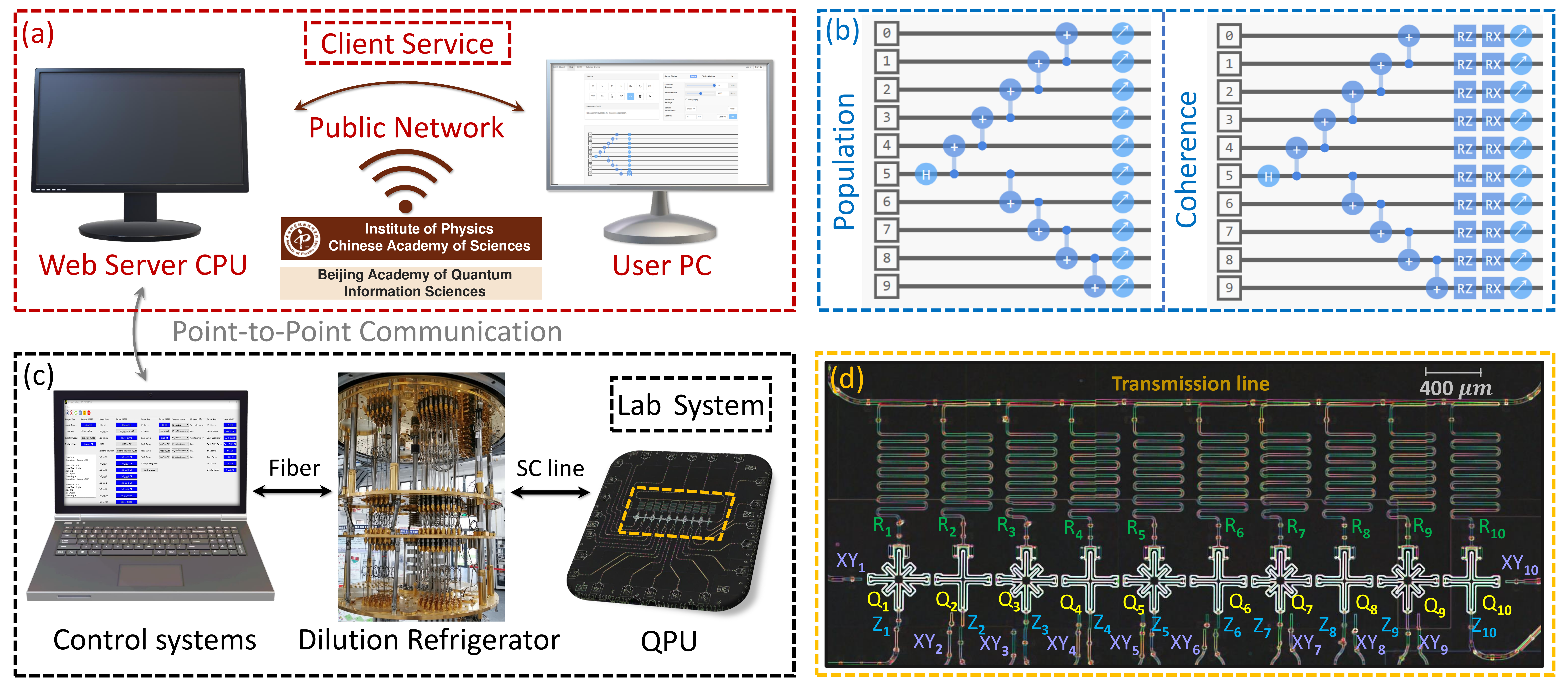}
\caption{(Color online) Framework of ScQ. (a) Client service of ScQ. Users can visit the ScQ website to construct the quantum circuit and run it. The web server central processing unit (CPU) of ScQ interacts with users through the public network and connects to the experiment control systems in the laboratory via point-to-point communication. (b) Quantum circuits of measuring the population (left) and coherence (right) of GHZ states. (c) Automatic control of running quantum algorithms on the superconducting quantum processing unit (QPU) in the lab system. On the experimental computer, we deploy an agent server as an intermediary to receive commands from the web. The commands of quantum operations will be compiled into microwave pulses and uploaded to the electronics corresponding to control lines of qubits. (d) 10-qubit chain QPU of ScQ. Each qubit can be independently controlled by the XY control line (microwave) and Z line (flux bias). Ten readout resonators are coupled to the qubits, which are connected to the transmission line for quantum non-demolition (QND) measurement~\cite{RMP_QND}. The QPU is mounted in a dilution refrigerator with the base temperature of the mixing chamber being approximately 10 mK.}
\label{framework}
\end{figure*}

Performing tasks with CQC is different from that in laboratories. For example, online gate fidelities will be generally lower because we can optimize the operation accuracy in the laboratory for a fixed scheme, which may not be applicable for CQC. In this sense, online gate fidelities can only be optimized for general purposes. CQC is designed to efficiently produce highly uniform outputs. We also strive to make it easy to operate. By exploiting CQC, the generation and particularly verification of multi-qubit GHZ states of up to 10 qubits remain challenging, even if it is a widely used performance benchmark for various platforms of quantum computation.

{\it Setup of ScQ.}---To set up the whole CQC platform, we allow users to access the web page of ScQ through a public network. Point-to-point communication is constructed between the lab computer and backend server for the web page. The framework of ScQ is summarized in Fig.~\ref{framework}. First, the tasks of quantum circuits submitted by different online users are sent to a web server for syntax checking and task scheduling. Then, the tasks are translated to certified order strings and subsequently sent to the lab system. An agent service program is installed in the lab computer for constantly requesting tasks from the web server. Once a task is received, it will be performed by the agent via the lab system. After the computing of the quantum processing unit (QPU), the results are pretreated by the lab computer and returned to the web server through the agent. Finally, the results are visually displayed on the web and users can download the results for individual data processing.

In the lab system, the QPU of ScQ consists of 10 qubits in a 1D array (see Fig.~\ref{framework}(d)~\cite{RMP_QND}). Considering the rotating wave approximation, the Hamiltonian of this system can be written as
\begin{equation}
	H/\hbar =\sum_{i}  -\frac{1}{2}\omega_{i}\sigma^{z}_{i} + 
	\sum_{i<j} g_{ij}(\sigma^{+}_{i}\sigma^{-}_{i+1}+h.c.)
\end{equation}
where $\omega_i$ is the frequency of qubit $Q_i$, $g_{ij}$ is the coupling coefficient of qubits, and $\sigma^{\pm}=(\sigma ^x\pm i\sigma ^y)/2$ is the raising (lowering) operator with $\sigma^{x,y,z}$ being Pauli matrices. The frequency of each qubit can be adjusted from 4.0 to 5.8 GHz. The NN qubits are directly coupled through capacitance, and the coupling strength is between 10 and 12 MHz. The coupling strength of the next NN qubits is one order of magnitude smaller than that of the NN qubits, which is approximately 1 MHz. All 10 qubits are coupled to one readout line through their individual readout resonators, the frequencies of which are between 6.49 and 6.66 GHz.

\begin{figure*}[ht]
	\
	\includegraphics[width=1\textwidth]{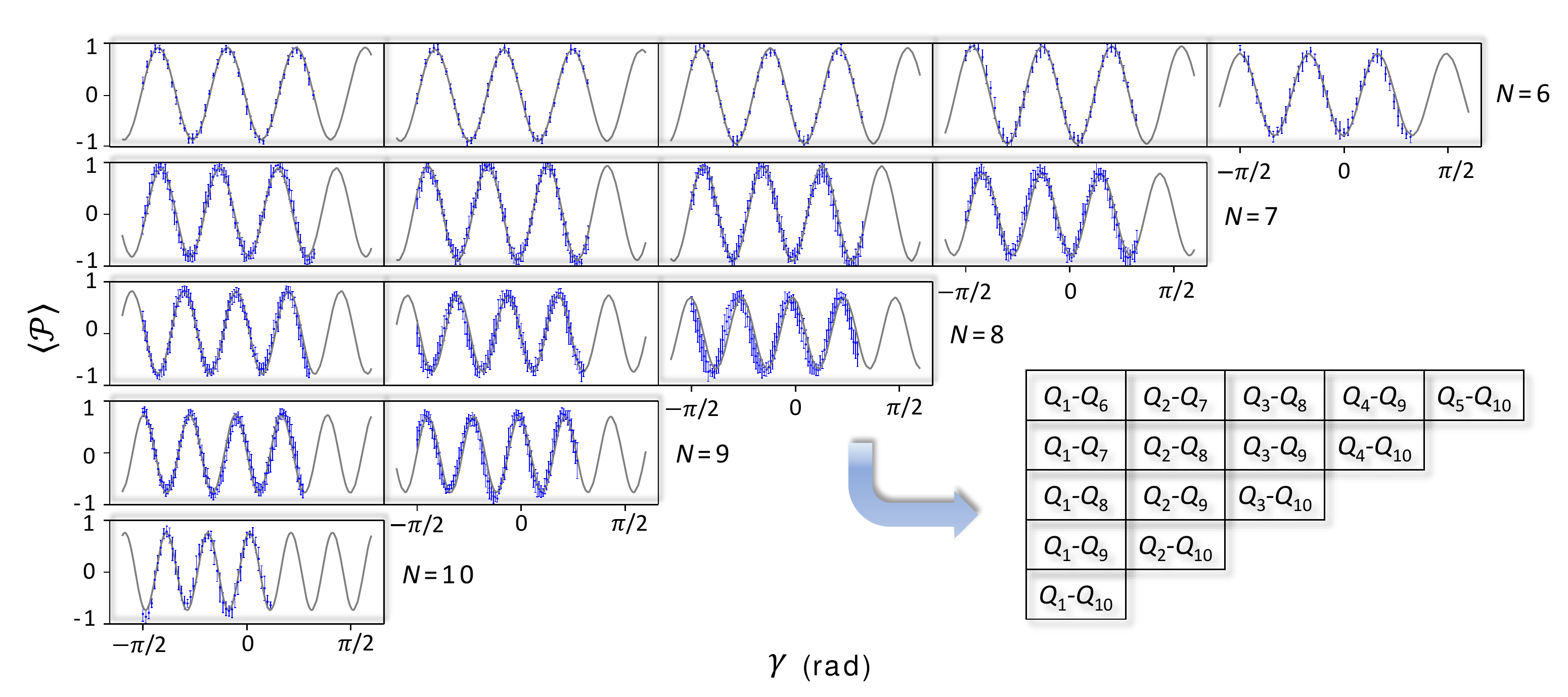}
	\caption{(Color online) Parity oscillations GHZ states with 6-10 qubits. Each row corresponds to GHZ states with the same number of qubits. From top to bottom, the qubits number of GHZ state are $N=6,~7,~8,~9,~10$. From left to right, the GHZ state consists of the first $N$ qubits and sequential $N$ qubits, respectively.}
	\label{parity_oscillations}
\end{figure*} 

In the calibration procedure, we bias each qubit to its idle frequency using an independent Z control line. Taking into account the AC Stark effect between qubits, we diverge the idle frequencies of NN qubits by at least 418 MHz and the frequency gap is even larger during the readout process. Hence, the crosstalk can be suppressed when qubits are manipulated and measured. Based on this, we calibrate single- and two-qubit gates in the lab system. The pulse durations of single-qubit gates are uniformly set to 30 ns, and the two-qubit CZ gate is around 40-ns-long. The average randomized benchmarking (RB)~\cite{RB1,RB2,RB3} fidelity of single-qubit gates reaches 99.7\%, whereas the average quantum process tomography (QPT)~\cite{QPT1,QPT2} fidelity of two-qubit CZ gates reaches 95.5\% (see Supporting Information).

All the above quantum gate operations can be remotely realized through the ScQ platform. Users can design their own quantum circuits using the drag-and-drop quantum gates toolbox of the graphic interface or with a quantum assembly (QASM) language of ScQ (\url{http://q.iphy.ac.cn}, 
 \url{http://quafu.baqis.ac.cn}). Particularly, ScQ allows users to scan the arbitrary parameters of single-qubit rotation gates. By clicking the `Parameter Setting' button of a quantum gate on the graphic interface, one can determine whether to use the gate with a fixed rotation angle in general setting or with varying parameters in an  advanced setting. ScQ also provides a QASM of coding in Python, which is similar to the Qiskit of IBM (\url{http://quantum-computing.ibm.com}). For instance, one can add quantum gates to manipulate qubits after initializing a \verb+QuantumCircuit+ object. In the following code, we show the basic imports and generation of a \verb+QuantumCircuit+ with 10 qubits: 
\begin{framed}
\begin{lstlisting}
import numpy as np
from numpy import pi
from qasm import QuantumCircuit
q = QuantumCircuit(10)
\end{lstlisting}
\end{framed}
\noindent
Here one can add gates to manipulate the qubits of \verb+QuantumCircuit+, such as adding a $R_x$ gate (\verb+rx+) with a fixed angle $\pi/2$ on the qubit 0:
\begin{framed}
\begin{lstlisting}
q.rx(0, pi/2)
q.measure([0])
result = q.send()
\end{lstlisting}
\end{framed}
\noindent
where the second line specifies the list of measured qubits, and the last line indicates that the task is sent to QPU and measurement result returns.

Compared with other CQC platforms, our ScQ provides a more direct way to express the `for' loop. For example, a $R_x$ gate with varying angle from $-\pi/2$ to $\pi/2$ on the qubit 0 can be directly expressed as
\begin{framed}
\begin{lstlisting}
angles = np.linspace(-pi/2, pi/2, 51)
q.rx(0, angles)
\end{lstlisting}
\end{framed}
\noindent
where the second parameter of \verb+rx+ gate is a list with multiple angles. This `for' loop of rotation is completed by the built-in loop in the lab computer connected to the experimental QPU. In this way, the time cost of messaging tasks and data processing will be reduced, especially in tasks concerning the measurement of off-diagonal elements.

{\it Online generation and verification of GHZ states.}---It is a performance standard for a quantum setup to produce multi-qubit entangled states with high fidelity. On the quantum processor of ScQ, we show the generation of high-fidelity GHZ states. Particularly, the GHZ states of all possible blocks of 1D array qubits from 6 to 10 qubits are prepared and verified, which provides a full figure of merit of the platform.   
The gate sequence for preparing a 10-qubit GHZ state is shown in Fig.~\ref{framework}(b), which consists of a Hadamard gate and subsequent series of CNOT gates. Each CNOT gate is composed of a lab-native CZ gate and two single-qubit rotations. Note that different sequences for preparing GHZ states have different layers of CNOT gates. We use the circuit in Fig.~\ref{framework}(b) to reduce the layers and crosstalk caused by parallel operations.

\begin{figure*}[ht]
	\
	\includegraphics[width=1.0\textwidth]{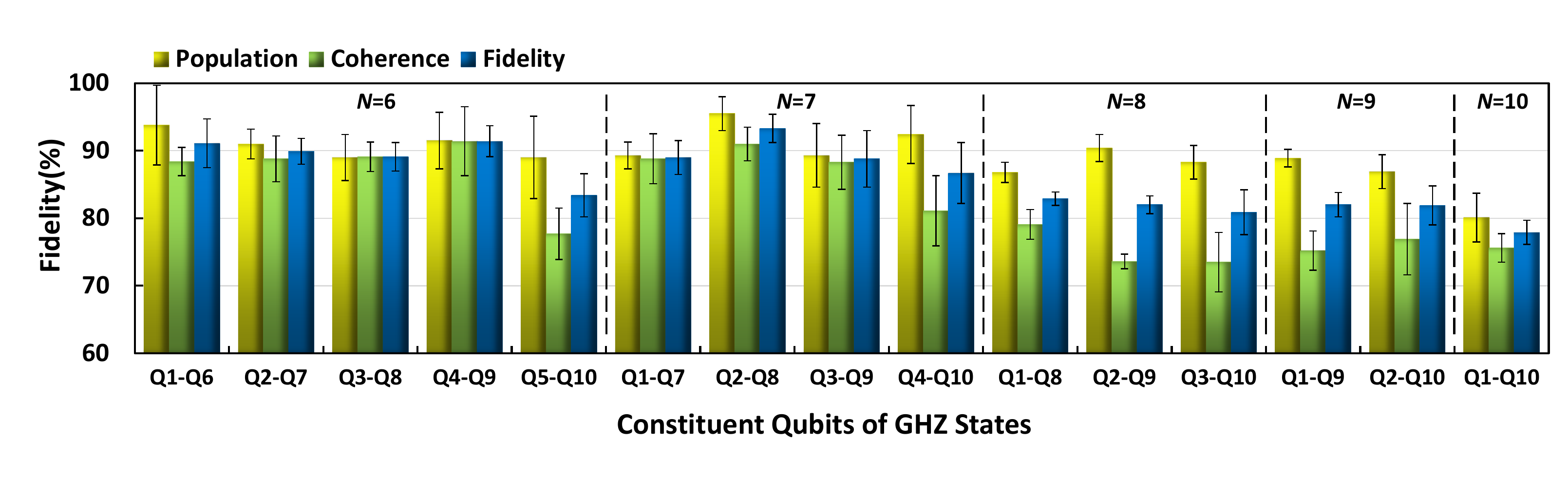}
	\caption{(Color online) Population, coherence,and fidelity of GHZ states with 6-10 qubits. From left to right, the qubit number of GHZ states are $N=6,~7,~8,~9,~10$.}
	\label{fidelity_GHZ}
\end{figure*}

In the QASM-like expression, the generation and characterization of GHZ states corresponding to Fig.~\ref{framework}(b) can be written as
\begin{framed}
\begin{lstlisting}
def ghz(n, angles = []):
    q = QuantumCircuit(n)
    ini = n // 2
    q.h(ini)
    for i in range(ini, 0, -1):
        q.cnot(i, i - 1)
    for i in range(ini, n - 1):
        q.cnot(i, i + 1)
    if angles:
        for i in range(n):
            a = [- j for j in angles]
	    q.rz(i, a)
	    for i in range(n):
	        q.rx(i, pi/2)
    q.measure([i for i in range(n)])
    res = q.send()
    return res

num = 10
angles = np.linspace(-pi/2, pi/2, 51)
results = ghz(num, angles)
\end{lstlisting}
\end{framed}
The above QASM code shows the characterization of the fidelity of 10-qubit GHZ states. The result of each loop is provisionally recorded in the lab computer and all results will return to the web once the whole experiment is completed.

Here, the fidelity of the generated GHZ state is defined as the distance between the prepared state and target state. The fidelity of a generated GHZ state should be above 0.5 \cite{PhysicsReport}, which ensures the genuine multi-qubit entanglement. We use the standard method to determine the fidelity, $F = (C + P)/2$, where $P$ represents the summation of populations of states $\ket{0...0}$ and $\ket{1...1}$ corresponding to two diagonal elements of the density matrix and $C$ denotes two off-diagonal elements corresponding to the relative coherence. We can obtain these two parameters experimentally. First, we obtain the population $P$ by making multiple measurements along the Z-axis and calculating the probabilities that the measurement results are in $\ket{0...0}$ state and $\ket{1...1}$ state, marked as $P(\ket{0...0})$, $P(\ket{1...1})$ respectively. Then we will get $P=P(\ket{0...0})+P(\ket{1...1})$.

In order to infer the relative coherence $C$, we need to introduce a rotation operator ${\cal{P}}(\gamma)= \bigotimes^{N}_{j=1}(\cos{\gamma}\sigma_{y,j}+\sin{\gamma}\sigma_{x,j})$~\cite{Monz,SongC17,SongC19,LukinScience}. The corresponding parity, written as ${\cal{P}}=C\cos{(N\gamma+\psi)}$ will oscillate by varying $\gamma$. It is clear that the relative coherence $C$ corresponds to the amplitude of ${\cal{P}}$. Moreover, the oscillation pattern of the parity ${\cal{P}}$ depends on number of qubits $N$ for the generated GHZ states. In the experiment, after the preparation of GHZ states, all the qubits should be rotated by the operation ${\cal{P}}(\gamma)$. Then, a Z-axis measurement will be applied, and the parity will be calculated ${\cal{P}} = {\cal{P}}_{even}-{\cal{P}}_{odd}$, where ${\cal{P}}_{even}$ and ${\cal{P}}_{odd}$ correspond to the summation of all the probabilities of states with an even number of qubits and an odd number of qubits in the state $\ket{1}$, respectively. With the change in $\gamma$, which is realized by scanning the parameter feature of ScQ, we will obtain an oscillation curve with $\gamma$. The period of the oscillation curve is related to the number of qubits of GHZ states. By fitting the experimental ${\cal{P}}$ with the cosine function, we will obtain the corresponding relative coherence $C$ of GHZ states. We prepared GHZ States with qubit number from 6 to 10 with different combinations of qubits for all possible blocks in the 1D array qubits. The parity oscillation curve corresponding to each GHZ state is shown in Fig.~\ref{parity_oscillations} and the results of fidelities are shown in Fig.~\ref{fidelity_GHZ}. The fidelity of 10-qubit GHZ state generated online reached 77.05\%, which approaches the best records achieved in labs~\cite{SongC19}, ranging from 66.80\% to 81.70\% for the 10-qubit case~\cite{IBMGHZ,GongM19,SongC17,SongC19}. The gate sequences prepared for the GHZ states with different qubit numbers are shown in Supplementary.

{\it Summary.}---In summary, we built an online CQC platform ScQ with 10 individually addressable superconducting qubits. The programmable QPU connected to the public network is available for every user to program arbitrary quantum circuits. Based on this device, we show how to use ScQ to generate and verify GHZ states. The preparation of GHZ states were performed on all possible blocks of qubits in the chain to provide a full evaluation of the performance of this platform. To facilitate a user-friendly experience, we allow users to manipulate qubits graphically or with a QASM programming language. The high controllability and efficiency of ScQ demonstrate the great potential of cloud computing architecture for creating new junctures in running quantum algorithms and studying quantum many-body problems. Although the results in current 10-qubit system can be obtained faster and of better quality on a classical computer with numerical simulations by software such as QuTiP~\cite{qutip2012}, the experimental realization will be necessary when the number of qubits increases. It can be expected that a more advanced device with more qubits and higher control accuracy will facilitate the applications of quantum computation.

\section*{Acknowledgements}
This work is supported by the Synergic Extreme Condition User Facility, National Natural Science Foundation of China (Grant Nos. T2121001, 11934018, 11904393, and 92065114), Strategic Priority Research Program of Chinese Academy of Sciences (Grant No. XDB28000000), Beijing Natural Science Foundation (Grant No. Z200009), Scientific Instrument Developing Project of Chinese Academy of Sciences (Grant No. YJKYYQ20200041), and the Key-Area Research and Development Program of Guangdong Province (Grant No. 2020B0303030001).


\clearpage
\widetext
\section{Supplemental materials: Cloud quantum computation for generating Greenberger-Horne-Zeilinger states of up to 10 qubits}

\section{Device information}
\begin{figure*}[ht]
\centering
\includegraphics[width=0.75\textwidth]{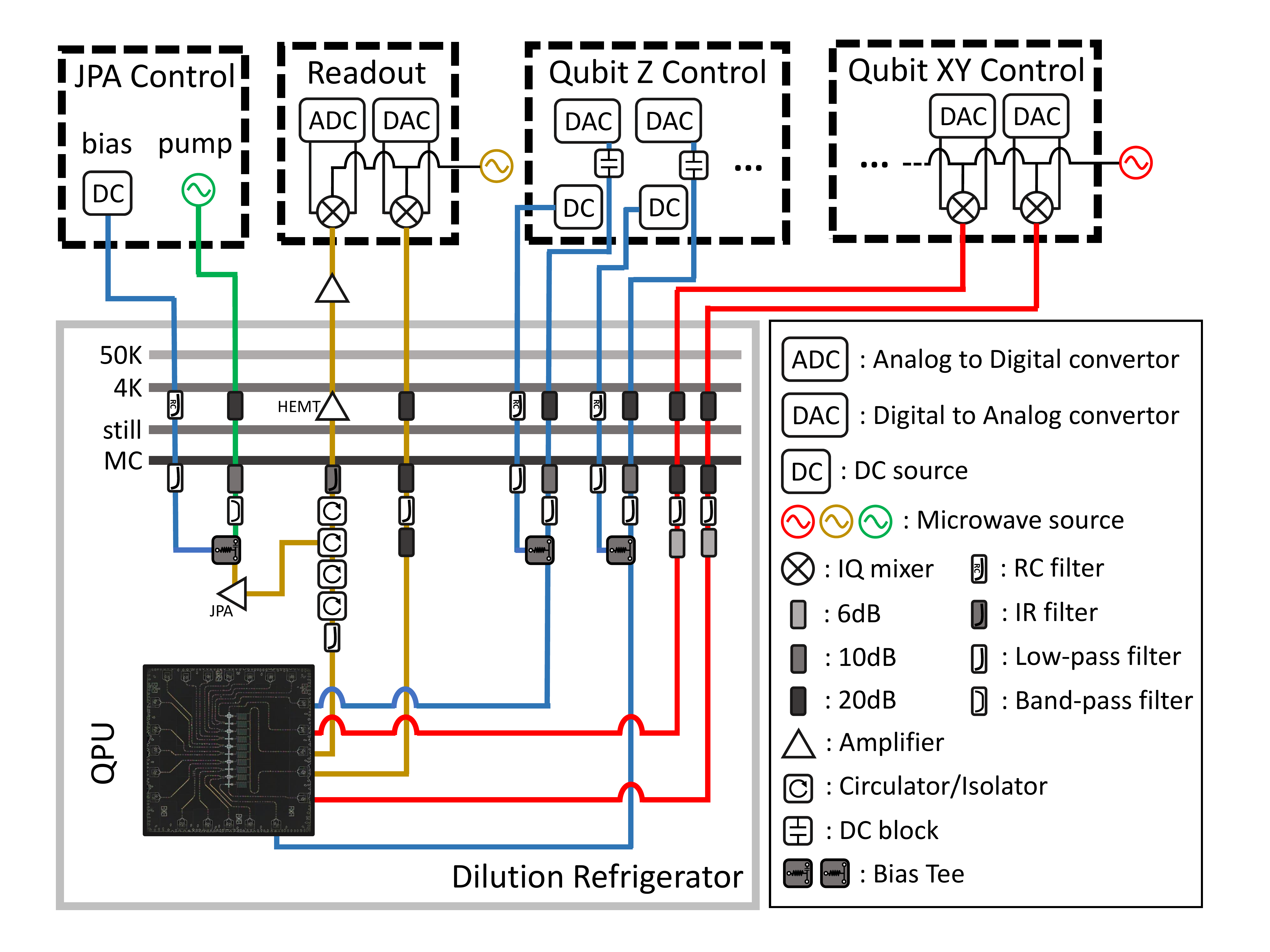}
\caption{Wiring of control electronics and cryogenic equipment. The QPU is
installed at the base temperature stage of the dilution fridge. We attach a 7.5 GHz, 500 MHz and 80 MHz low-pass filter to the XY control line, fast Z control line and DC Z control line, respectively. For Z control, a bias tee is used to couple the fast and DC Z control line. In order to enhance the signal noise ratio (SNR) of readout signal, we use a Josephson parametric amplifier (JPA) to amplify the readout signal by reflection, which is powered by a high-frequency pump and biased by a DC source. All the qubits can be measured simultaneously.}
\label{wiring}
\end{figure*}

Our quantum processor consists of 10 transmon qubits ($Q_1-Q_{10}$) arrayed in a row, where each qubit is capacitively coupled to its nearest-neighbors with the coupling magnitude listed in Table.~\ref{para}. The whole wiring of control electronics and cryogenic equipment are shown in Fig.~\ref{wiring}. Each qubit is coupled to its own readout resonator, which can be probed though a common transmission line connected to them for qubit measurement. Each transmon qubit can be individually addressed through its flux line for dynamical frequency modulation and microwave drive line for state excitation. The qubit parameters and coherence performance can be found in Table.~\ref{para}. In experiment, the frequencies of qubits is tunable from about 4 to 5.5 GHz. We design the idle frequencies $\omega_{10}^j$ of each qubit to avoid crosstalk of single-qubit gates. The interacting frequencies $\omega^I_{j,j+1}$ ($j$ from 1$-$9) of performing two-qubit gates are different in order to reduce the effects between qubits when quantum gates are excuted 
in parallel. Note that coupling strengths between adjacent qubits range from 10 to 12 MHz, which ensures that the two-qubit control-Z gate can be completed within 50 ns with the non-adiabatic scheme adopted in our experiment.


\begin{table}[h]
    \centering
    \setlength{\tabcolsep}{3mm}{
    \begin{tabular}{|ccccccccccc|} 
        \hline qubit&$Q_{1}$&$Q_{2}$&$Q_{3}$&$Q_{4}$&$Q_{5}$&$Q_{6}$&$Q_{7}$&$Q_{8}$&$Q_{9}$&$Q_{10}$\\
        \hline $\omega^{s}_{j}/2\pi$ (GHz)  &5.536&5.069 &5.660 &4.742 &5.528 &4.929 &5.451 &4.920 &5.540 &4.960 \\
        $\omega_{10,j}/2\pi$  (GHz)    &5.492 &4.843 &5.420 &4.780 &5.468 &4.737 &5.384 &4.880 &5.298 &4.714 \\
        $\omega^{r}_{j}/2\pi$  (GHz) & 5.493 &4.800 &5.455 &4.734 &5.300 &4.436 &5.252 &4.807 &5.309 &4.402 \\
        $\eta_{j}/2\pi$   (GHz) & 0.250 &0.207 &0.251 &0.206 &0.251 &0.203 &0.252 &0.204 &0.246 &0.208 \\
        $g_{j,j+1}/2\pi$ (MHz) & 12.07 &11.58 &10.92 &10.84 &11.56 &10.00 &11.74 &11.70 &11.69 & - \\
        $\omega^{I}_{j,j+1}/2\pi$ (GHz) & 5.152 & 4.843 & 5.071 & 4.780 & 5.033 & 4.737 & 5.045 & 4.880 & 4.973 & -\\
        $\omega^{I}_{j-1,j}/2\pi$ (GHz) & - & 4.902 & 5.094 & 4.820 & 5.031 & 4.782 & 4.989 & 4.793 & 5.126 & 4.726\\
        $T_{1,j}$ (us)    &23.0 &25.1 &42.5 &29.7 &27.51 &32.1 &23.7 &52.3 &25.7 &24.6 \\
        $T_{2,j}^{*}$ (us)    &22.3 &1.92 &4.23 &2.94 &9.99 &2.43 &11.17 &2.82 &3.49 &2.31 \\
        $F_{0,j}$ (\%) & 98.50 &98.45 &97.40 &98.13 &97.47 &96.43 &96.60 &96.37 &98.17 &98.00 \\
        $F_{1,j}$ (\%) & 94.20 &93.97 &94.80 &94.27 &92.03 &86.80 &93.03 &90.97 &92.17 &88.30 \\

        $X$ fidelity (\%)&99.95 &99.90 &99.84 &99.94 &99.68 &99.85 &99.83 &99.73 &99.82 & 99.89 \\
        $X$/2 fidelity (\%)&99.79 &99.85 &99.82 &99.83 &99.78 &99.60 &99.60 &99.88 &99.62 & 99.78 \\
        \hline
        
    \end{tabular}}
    \caption{Device parameters. $\omega^{s}_{j}$ represents the maximum frequency of $Q_{j}$. $\omega_{10,j}$ corresponds to the idle frequency of $Q_{j}$. $\omega^{r}_{j}$ indicates the resonant frequency of $Q_{j}$ during readout. $\eta_j$ shows the anharmonicity of $Q_{j}$. $g_{j,j+1}$ is the coupling strength between nearest-neighbor qubits. $\omega^{I}_{j,j+1}$($\omega^{I}_{j-1,j}$) represents the interaction point for generating the CZ gate composed of $Q_{j},Q_{j+1}$($Q_{j-1},Q_{j}$). $T_{1,j}$ and $T^{*}_{2,j}$ represent the relaxation time and coherence time of $Q_{j}$. $F_{0,j}$ and $F_{1,j}$ are readout fidelities of $Q_{j}$ in $\ket{0}$ and $\ket{1}$, which can be used for readout correction. The single-qubit $X(X/2)$-gate fidelity of is measured when all other qubits are at their idle points. 
    }
    \label{para}
\end{table}

\section{Calibration and optimization of two-qubit gates}


In experiment, we adopt an non-adiabatic scheme to realize the two-qubit CZ gate between $Q_j$ and $Q_{j+1}$. We apply the rectangular pulses to each qubit simultaneously on their flux bias lines, tuning qubit frequencies non-adiabatically to make $\omega_{10}$ of $Q_j$ to be resonant with the $\omega_{21}$ of $Q_{j+1}$ for a specific time $t$, thus realizing a $\vert1_{j}1_{j+1}\rangle \Leftrightarrow \vert0_{j}2_{j+1}\rangle$ swap process. By finely adjust $t$, we can maximize the probability of the final state in $\ket{11}$, then we realize a CZ gate between $Q_j$ and $Q_{j+1}$. In experiment, the rectangular pulse amplitude for each $Q_j$ is optimized by a ramsey-like sequence \cite{Xu2020, Li2020, Han2020, Foxen2020}, aiming to make the accumulated phase difference of the target qubit $Q_{j+1}$ between the case when control qubit $Q_j$ stays in $\vert 0\rangle$ and that when $Q_j$ stays in $\vert 1\rangle$ state is $\pi$ after the CZ sequences. The interaction time $t$ are selected when the initially prepared $\vert 1_{j}1_{j+1}\rangle$ state can be nearly brought back after experiencing a full $\vert1_{j}1_{j+1}\rangle\Leftrightarrow\vert0_{j}2_{j+1}\rangle$ swap process.

\begin{table}[ht]
    \centering
    \setlength{\tabcolsep}{2mm}{
    \begin{tabular}{|cccccccccc|} 
        \hline Constituent qubits & $Q_{1}Q_{2}$ & $Q_{2}Q_{3}$ & $Q_{3}Q_{4}$ & $Q_{4}Q_{5}$ & $Q_{5}Q_{6}$ & $Q_{6}Q_{7}$ & $Q_{7}Q_{8}$ & $Q_{8}Q_{9}$ & $Q_{9}Q_{10}$ \\
        \hline $T_{\mathrm{CZ}} (\mathrm{ns})$ & 48.9 & 46.1 & 40.1 & 39.0 & 42.8 & 41.8 & 43.6 & 40.5 & 41.1 \\
        $F_{\mathrm{ini}}$ (\%) & 98.58 & 98.42 & 98.66 & 98.98 & 98.76 & 98.70 & 97.90 & 97.52 & 98.68 \\
        $F_{\mathrm{fin}}$ (\%) &98.00 &97.42 &97.30 &97.88 &97.92 &95.66 &97.12 &96.48 &97.72 \\

        
        $F_{\chi}$ (\%) &96.82(1) &95.40(1) &94.70(1) &94.48(1) &94.72(1) &92.76(2) &94.82(1) &93.18(1) &95.84(1) \\
        \hline 

    \end{tabular}}
    \caption{Gate fidelities of CZ gates. $T_{\mathrm{CZ}}$ represents the total duration of the pulse forming CZ gate. $F_{\mathrm{ini}} (F_\mathrm{fin})$ is the mean fidelity of all initial(final) states in QPT process. $F_{\chi}$ is the process fidelity of CZ gate obtained through QPT process.}
    \label{CZ}
\end{table}

The fidelities of two-qubit CZ gates are characterized by quantum process tomography (QPT). In brief, we prepare 16 initial states of two qubits, namely $\left\{\ket{0}, \ket{1}, (\ket{0}+\ket{1})/\sqrt{2}, (\ket{0}-\mathrm{i}\ket{1})/\sqrt{2}\right\}^{\otimes2}$, and perform quantum state tomography (QST) after the experiment process. The initial state $\rho_{\mathrm{ini}}$ and final state $\rho_{\mathrm{final}}$ satisfy
\begin{equation}
\rho_{\mathrm{final}} = \sum_{n,m}\chi_{nm}E_n\rho_{\mathrm{ini}}E_m^{\dagger},
\end{equation}
where the experimental process matrix is $\chi$ and $E_n$ is so-called Kraus operator that is chosen from the set $\left\{I, \sigma_x, -\mathrm{i}\sigma_y, \sigma_z\right\}^{\otimes2}$. For the process of CZ gate, one can calculate the ideal $\chi_\mathrm{ideal}$ and measure $\chi_\mathrm{exp}$ of the experiment process. The process fidelity thus is given by $F_{\chi}=\mathrm{Tr}(\chi_\mathrm{exp}\chi_\mathrm{ideal})$. The optimization of CZ gate depends on the criterion of high-fidelity $\chi$ matrix, which accurately reflects the process of the interaction between two qubits.

The QPT fidelities of each qubit pair $\{Q_{j},Q_{j+1}\}$, shown in table.~\ref{CZ}, are obtained while other qubits are in their ground states at idle points. Note that the CZ fidelity estimated with QPT includes the state preparation error and measurement error. We did not utilize the RB (randomized benchmarking) method because when running a RB sequence including a large number of pulses, the rectangular pulse, which is used to bias the qubit to the idle point, needs to be very long. However, because of the imperfection of our arbitrary waveform generator (AWG), the long square pulse it generates become a little tilted, thus seriously affects the RB fidelity. The problem will be addressed in our second-generation experimental platform.

\section{Pulse shape optimization}
\begin{figure*}[ht]

\centering

\includegraphics[width=0.8\textwidth]{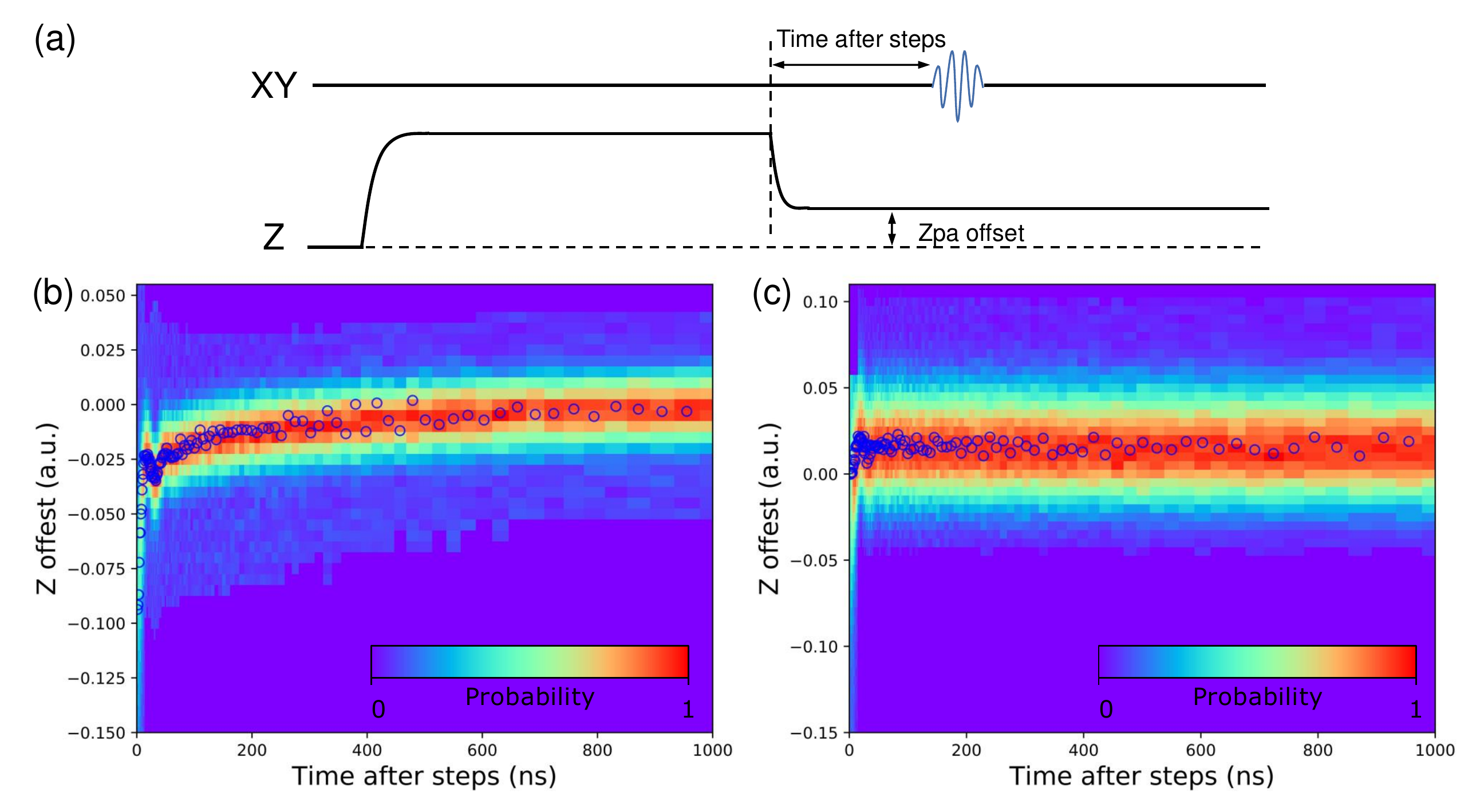}

\caption{Pulse sequence and data for measuring pulse shape. (a) displays the pulse sequence for measuring Z-bias pulse shape. (b) shows the response of qubits to step pulses without correction, and (c) shows the response with correction used. After correction, the pulse distortion is reduced.}

\label{pulseshape}
\end{figure*}

As mentioned above, the implementation of CZ gate requires two qubits to stay at the interacting frequency for a specific time $t$ through rectangular pulses imposed on their flux lines, The pulse distortion caused by the impedance mismatch on the flux line can have a serious effect on the fidelity of the CZ gate. Therefore, it is very important to correct this signal distortion. In our experiment, we perform an experimental sequence called Pulseshape to measure the response function of a step-edge, which can be used to correct the rectangular pulses, as can be seen in Fig.~\ref{pulseshape}(a). The qubit was biased to the flux-sensitive point, thus enabling us to accurately prob the frequency offset caused by the signal distortion. We then imposed a long square pulse on the flux line to create a step edge. The time-dependent qubit frequency shift caused by this step-edge was detected by applying a short Gaussian pulse with a FWHM of 10 ns to the qubit at different time $t$ after the square pulse. We use exponential(polynomial) function to fit the data after(before) 150 ns \cite{Rol2020}. The experimental results before and after correction are shown in Fig.~\ref{pulseshape}(b)(c).





\section{Crosstalk correction}
\begin{figure*}[h]
\centering
\includegraphics[width=0.6\textwidth]{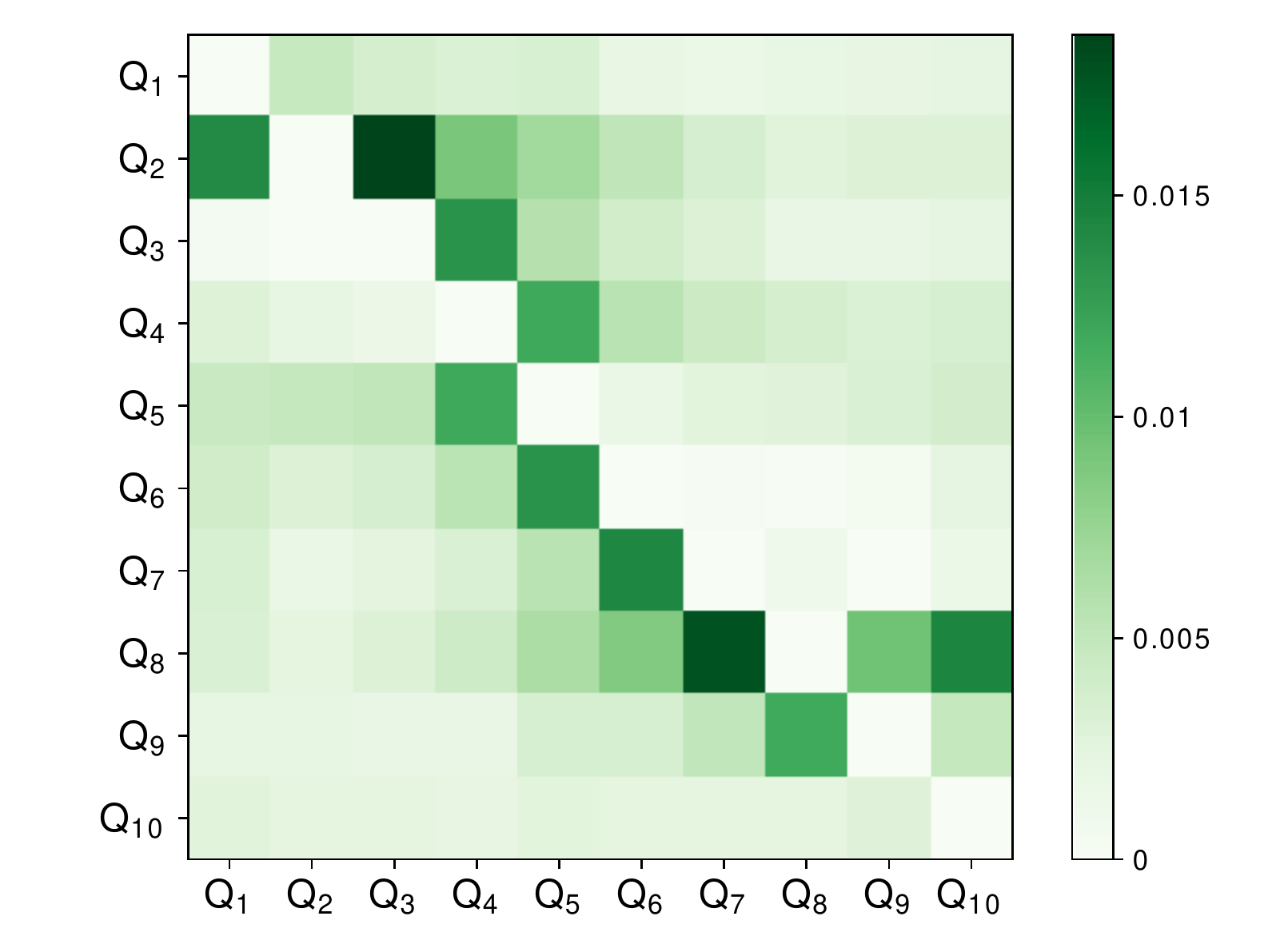}
\caption{Crosstalk matrix of the quantum processor. In the crosstalk matrix, the colour depth in the same row as $Q_{i}$ and in the same column as $Q_{j}$ indicates the ratio of Z-signal strength applied to $Q_{j}$ and $Q_{i}$ when the influence of Z-signal applied to $Q_{i}$ on $Q_{j}$ is the same as that of Z-signal applied to $Q_{j}$ itself. For clarity, we change the value of the point with $i = j$ in the image from 1 to 0.
}
\label{crosstalk}
\end{figure*}

In our experiment, there exists Z-crosstalk effects between qubits, which can be qualified by measuring the Z-crosstalk matrix $M^z$ (see Tab. \ref{Matrix}), where element $M^z_{i,j}$ represents the Z bias of $Q_j$ senses due to a unitary bias applied to $Q_i$. $M^z_{i,j}$ can extracted from the following method. We place $Q_j$ about 1 GHz below its sweet point, where the qubit frequency is sensitive to the Z signal. Then a resonant fixed-frequency microwave pulse with a flattop envelop is applied, during which we impose a square pulse with a large amplitude on Z line of $Q_i$ and an offset pulse on Z line of $Q_j$ simultaneously to cancel the frequency deviation caused by the large rectangular pulse on Z line of $Q_i$ through crosstalk effect. The experimental results are shown in Fig.~\ref{crosstalk}, from which $M^z_{i,j}$ can be extracted by fitting the slope.

To correct the Z-crosstalk effects with the experimentally obtained $M^Z$, we just perform the following numerical calculation to the waveform data by $Z^{\mathrm{applied}}=(M^z)^{-1}Z^{\mathrm{actual}}$, where $Z^{\mathrm{applied}}$ is the amplitudes of the Z biases imposed on Z lines of qubits and $Z^{\mathrm{actual}}$ is the Z bias amplitudes the 10 qubits actually sense, which is written in a column format.

\begin{table}[ht]
	\begin{equation}
        \begin{pmatrix}
        1 & -0.005 & -0.004 & -0.003 & -0.003 & -0.002 & -0.002 & -0.002 & -0.002 & -0.002 \\
        0.014 & 1 & -0.019 & -0.009 & 0.007 & 0.005 & 0.003 & 0.003 & 0.003 & 0.003 \\
        0 & 0 & 1 & -0.013 & 0.006 & 0.004 & 0.003 & 0.002 & 0.002 & -0.002 \\
        -0.003 & 0.002 & 0.001 & 1 & 0.012 & 0.005 & 0.004 & 0.004 & 0.003 & -0.003 \\
        -0.005 & 0.005 & 0.005 & 0.012 & 1 & 0.002 & 0.003 & 0.003 & 0.003 & -0.004 \\
        -0.004 & 0.003 & 0.004 & 0.005 & -0.013 & 1 & 0 & 0 & 0 & -0.002\\
        -0.003 & 0.002 & 0.002 & 0.003 & -0.005 & -0.001 & 1 & -0.001 & 0 & -0.001 \\
        -0.003 & 0.002 & 0.003 & 0.004 & -0.006 & -0.009 & -0.018 & 1 & -0.009 & 0.014\\
        -0.002 & 0.002 & 0.002 & 0.002 & -0.003 & -0.003 & -0.005 & -0.012 & 1 & -0.005\\
        -0.003 & 0.002 & 0.002 & 0.002 & -0.003 & -0.002 & -0.002 & -0.002 & -0.03 & 1\\
        \end{pmatrix}
	\end{equation}
    \caption{Z-crosstalk matrix $M^z$. The matrix element $M^z_{i,j}$ represents the Z bias of $Q_{j}$ senses  due to a unitary bias applied to Z line of $Q_{i}$.}
    \label{Matrix}
\end{table}


\section{Readout correction}

Because of the readout error in the qubit measurement, the experimentally measured raw probabilities need to be corrected before further treatment to extract the real GHZ fidelity. We assume there is no readout crosstalk between qubits. In fact, the readout crosstalk can be nearly cancelled by finely tuning the qubit frequencies when applying microwave pulses to readout resonators to avoid frequency crossover between qubits caused by AC-stark shift of resonator photon pumping. In Table.~\ref{para}, we show the probability of reading out $Q_j$ in $\vert 0\rangle$ ($\vert 1\rangle$) when it is prepared in $\vert 0\rangle$ ($\vert 1\rangle$) state, denoted as  $F^{j}_{0}$ and  $F^{j}_{1}$ respectively. The readout correction matrix of $Q_j$ can be written as

\begin{equation}
M_{j}=
\left(
\begin{matrix}
F^{j}_{0}   &  1-F^{j}_{1}\\
1-F^{j}_{1} &  F^{j}_{1}
\end{matrix}
\right)
\end{equation}
    
If only $Q_j$ is measured separately, the uncorrected probabilities of $\ket{0}$ state and $\ket{1}$ state shall be written as
\begin{equation}
\left(
\begin{matrix}
p^{j,\mathrm{uncorrected}}_{0}  \\
p^{j,\mathrm{uncorrected}}_{1}
\end{matrix}
\right)
=M_{j}
\left(
\begin{matrix}
p^{j}_{0}  \\
p^{j}_{1}
\end{matrix}
\right)
=
\left(
\begin{matrix}
F^{j}_{0}p^{j}_{0} + (1-F^{j}_{1})p^{j}_{1}  \\
(1-F^{j}_{0})p^{j}_{0} + F^{j}_{1}p^{j}_{1}
\end{matrix}
\right)
\end{equation}
We have repeatedly adjusted the frequency arrangement when all the qubits are staying at the reading point, so that there is almost no reading crosstalk between qubits. Therefore, in the case of simultaneous reading of multiple qubits, the total correction matrix can be approximately written as the direct product of every single qubit correction matrix:
\begin{equation}
M = M_{0} \otimes M_{1} \otimes \cdots \otimes M_{N},
\end{equation}
where $N$ is the total number of measured qubits. Assume that measured probability of the qubits string before correction is $\tilde{\textbf{P}}^{\mathrm{uncorrected}}$, one can obtain the probability after correction as
\begin{equation}
{\textbf{P}^{\mathrm{corrected}}}=M^{-1} \tilde{\textbf{P}}^{\mathrm{uncorrected}}.
\end{equation} 
For generality, our ScQ provide both $\textbf{P}^{\mathrm{corrected}}$ and $\textbf{P}^{\mathrm{uncorrected}}$ to the users. Advanced correction can be performed by using the original probabilities if needed.

\section{Quantum circuits of the GHZ states}
\begin{figure}[ht]
\centering
\includegraphics[width=0.31\textwidth]{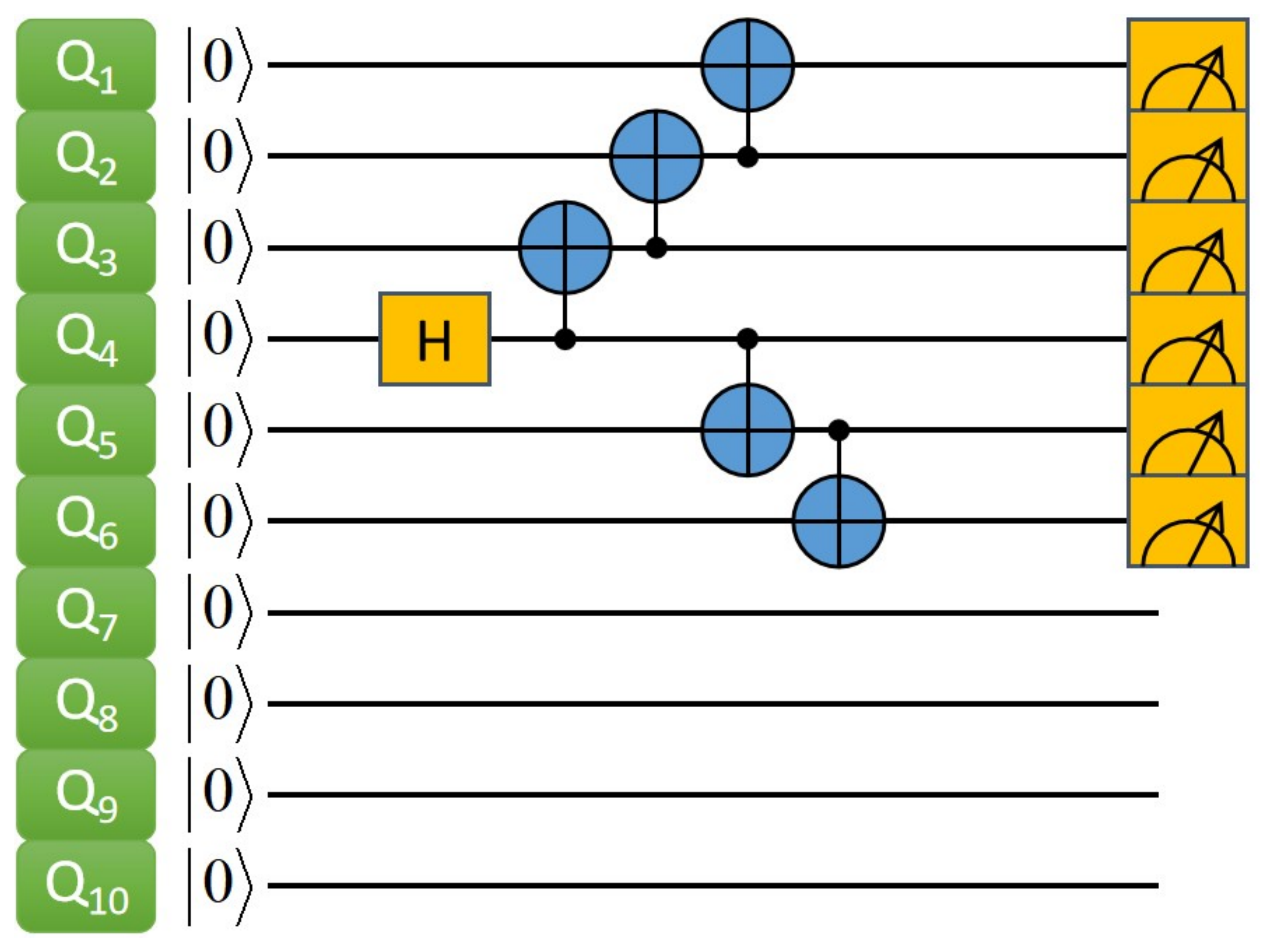}
~~
\includegraphics[width=0.31\textwidth]{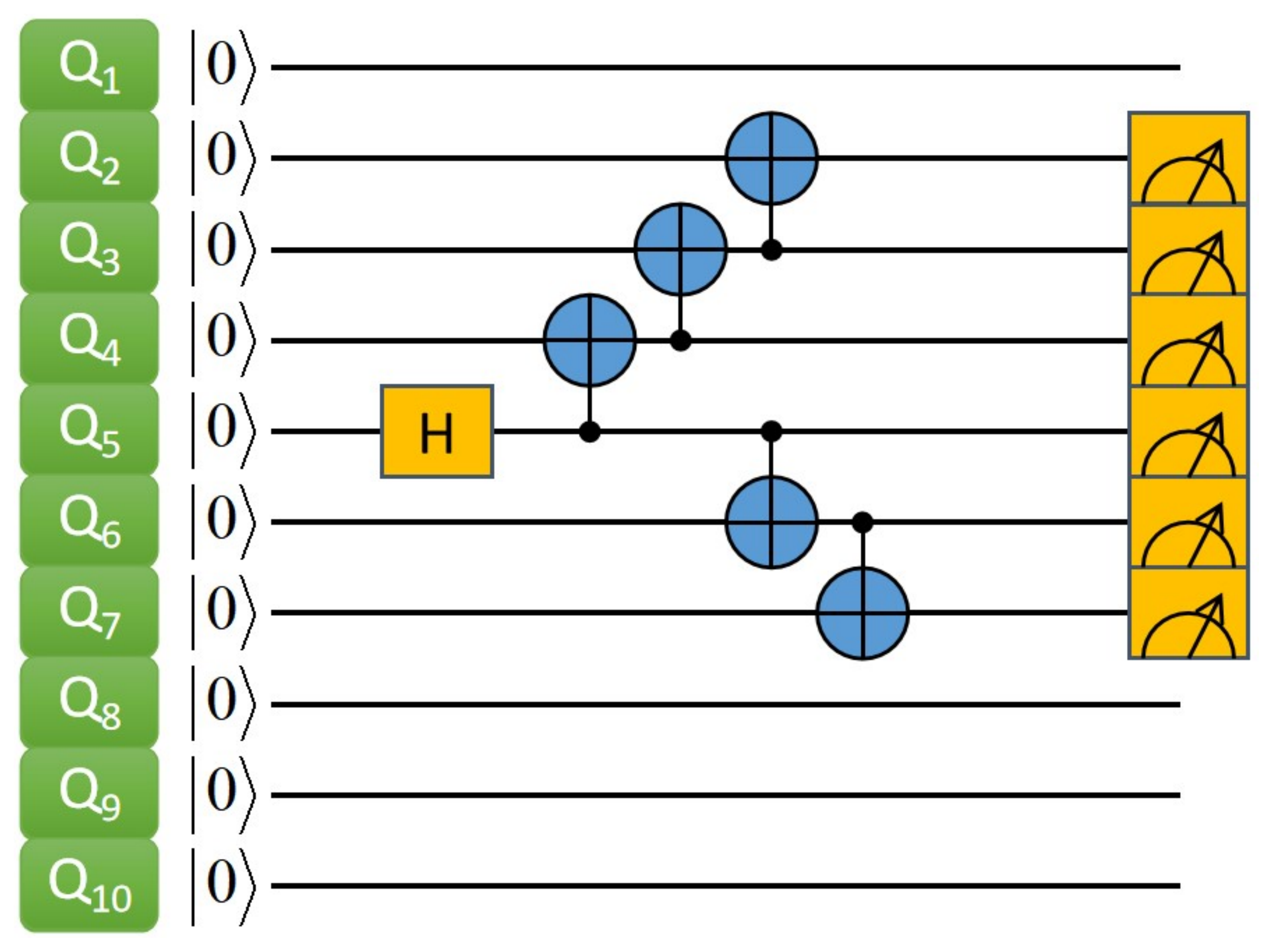}
~~
\includegraphics[width=0.31\textwidth]{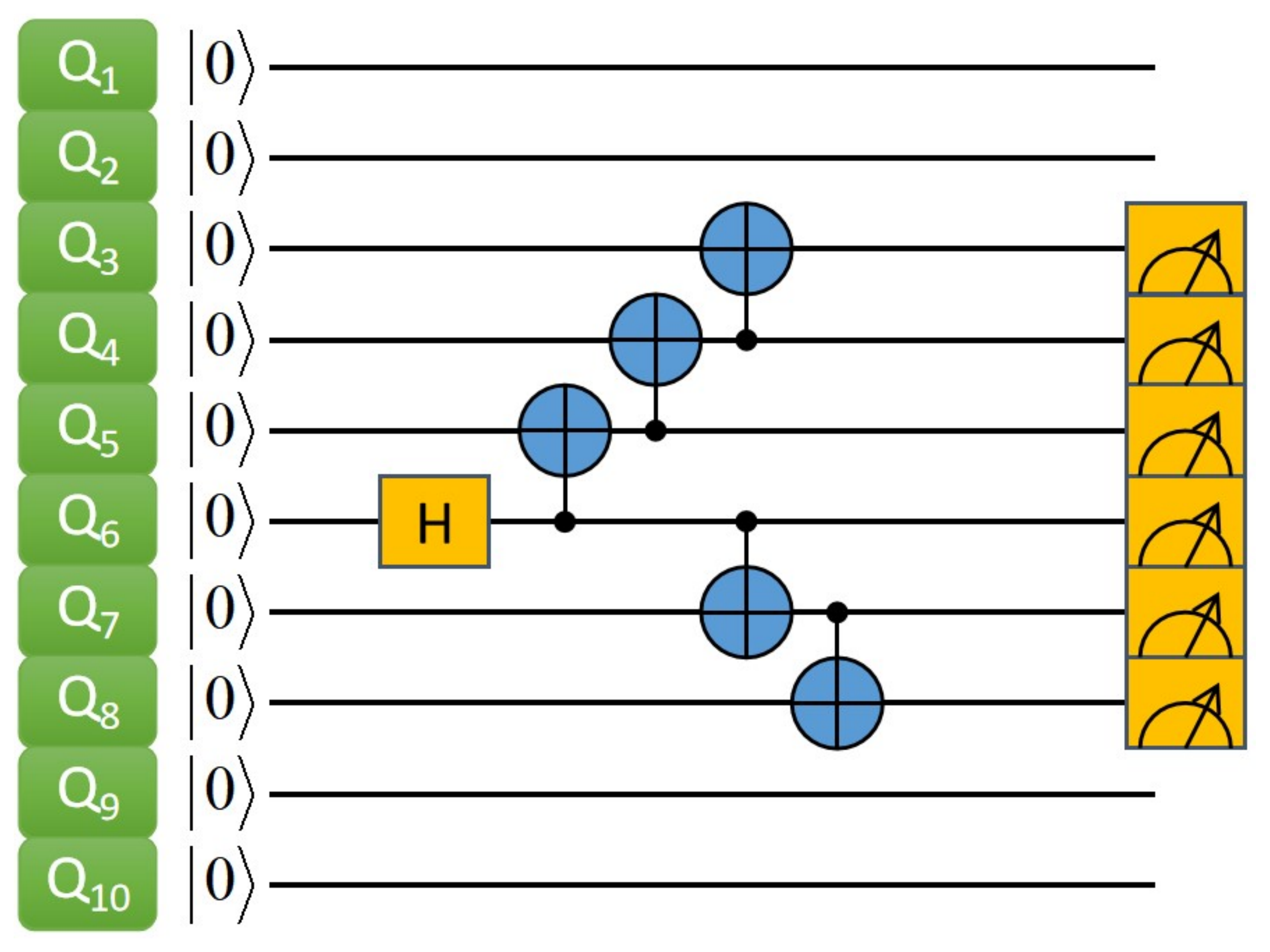}
\\
\vskip 0.5cm
~~~
\includegraphics[width=0.31\textwidth]{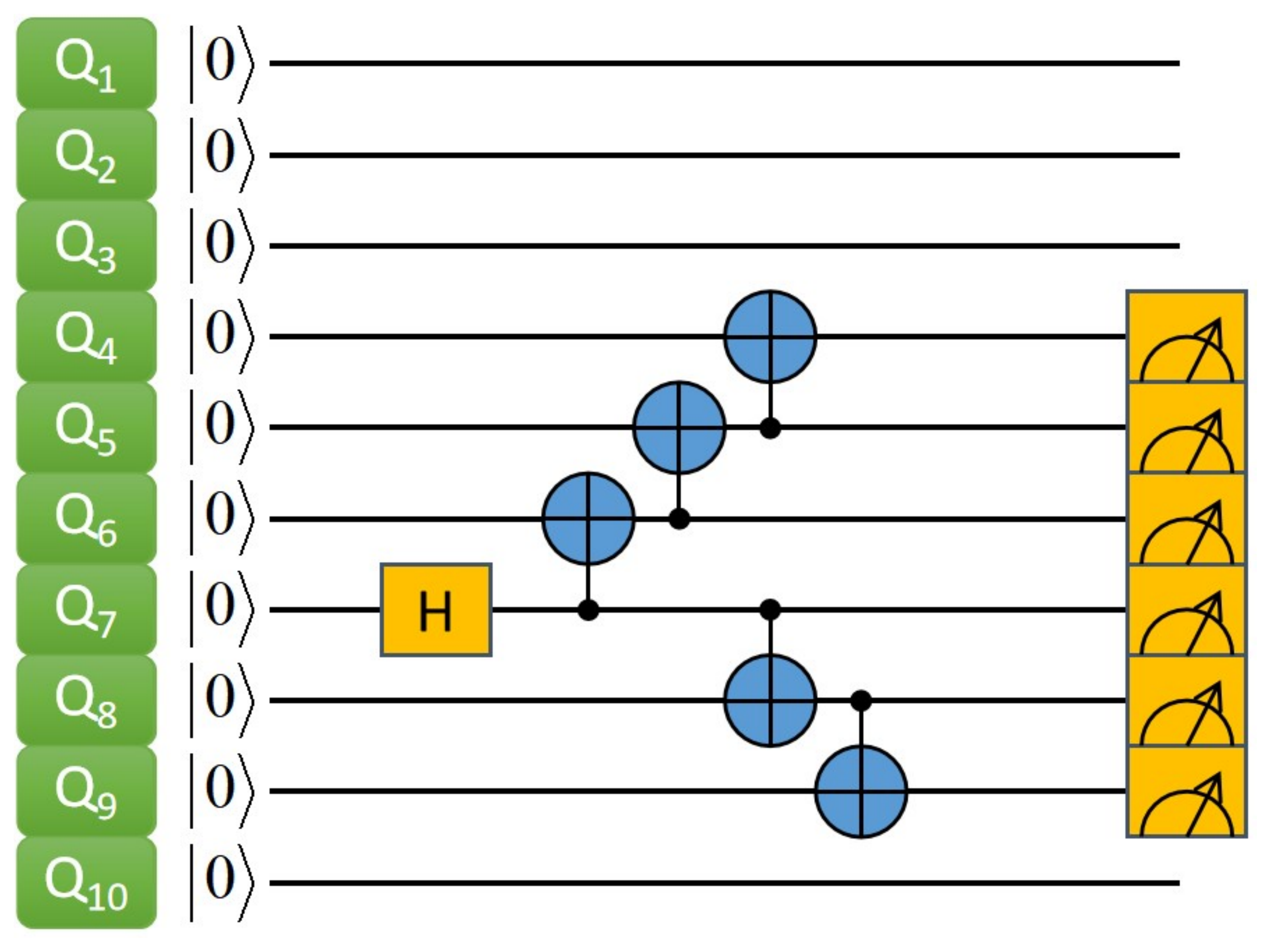}
~~~
\includegraphics[width=0.31\textwidth]{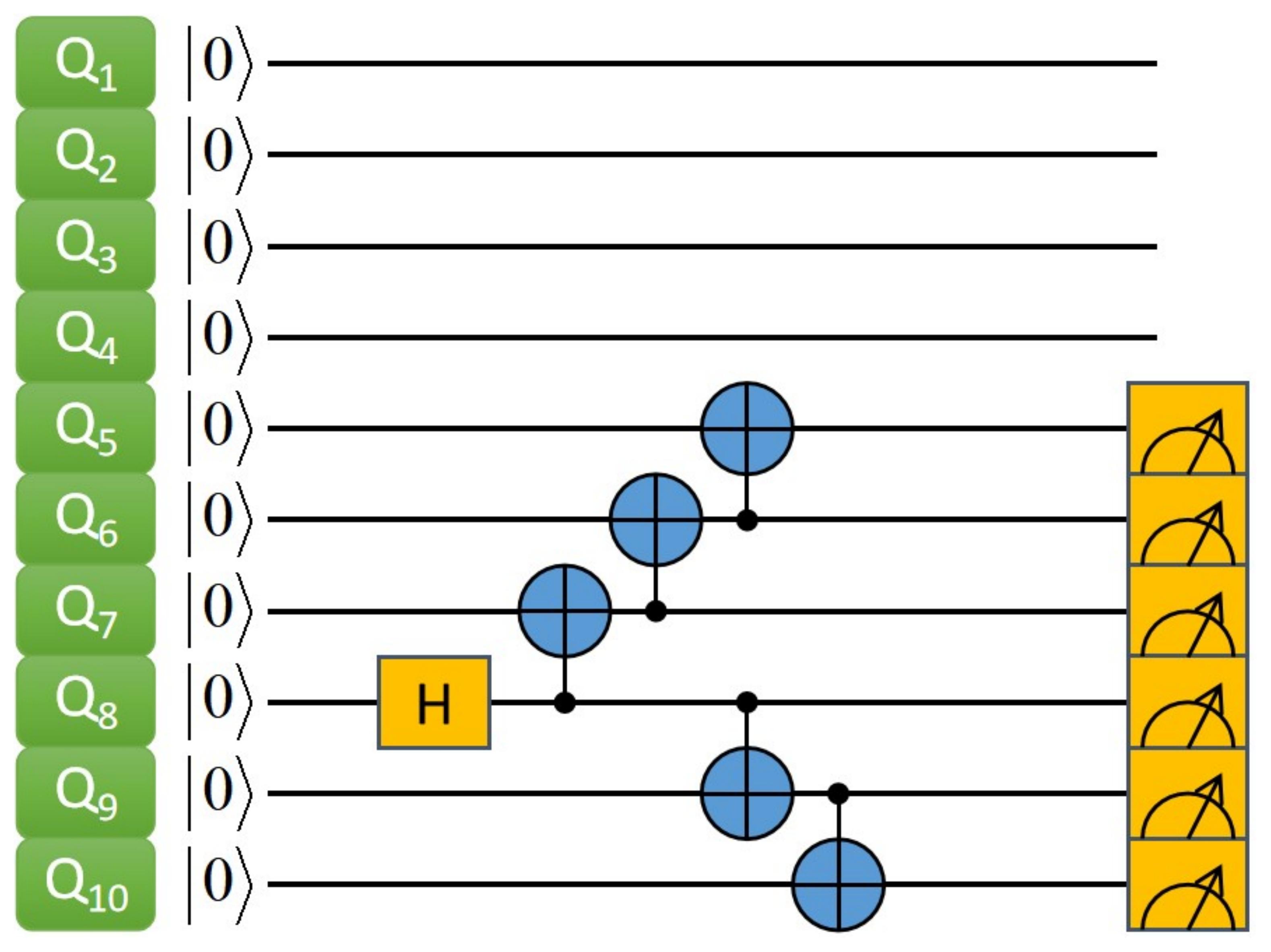}
\caption{Quantum circuits of 6-qubit GHZ state generation}
\label{6qubitGHZ}
\end{figure}

\begin{figure}[ht]
\centering
\includegraphics[width=0.31\textwidth]{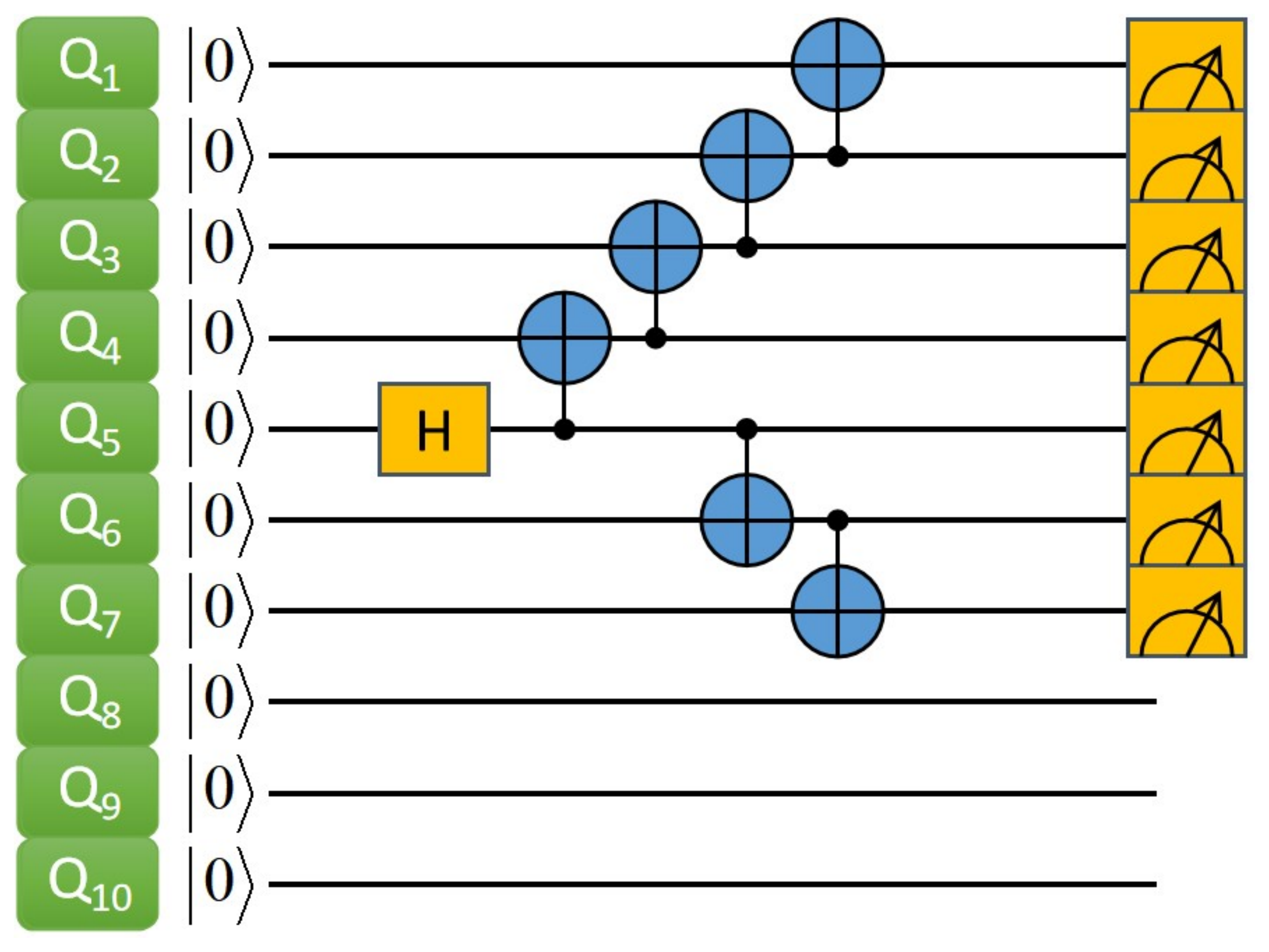}
~~~
\includegraphics[width=0.31\textwidth]{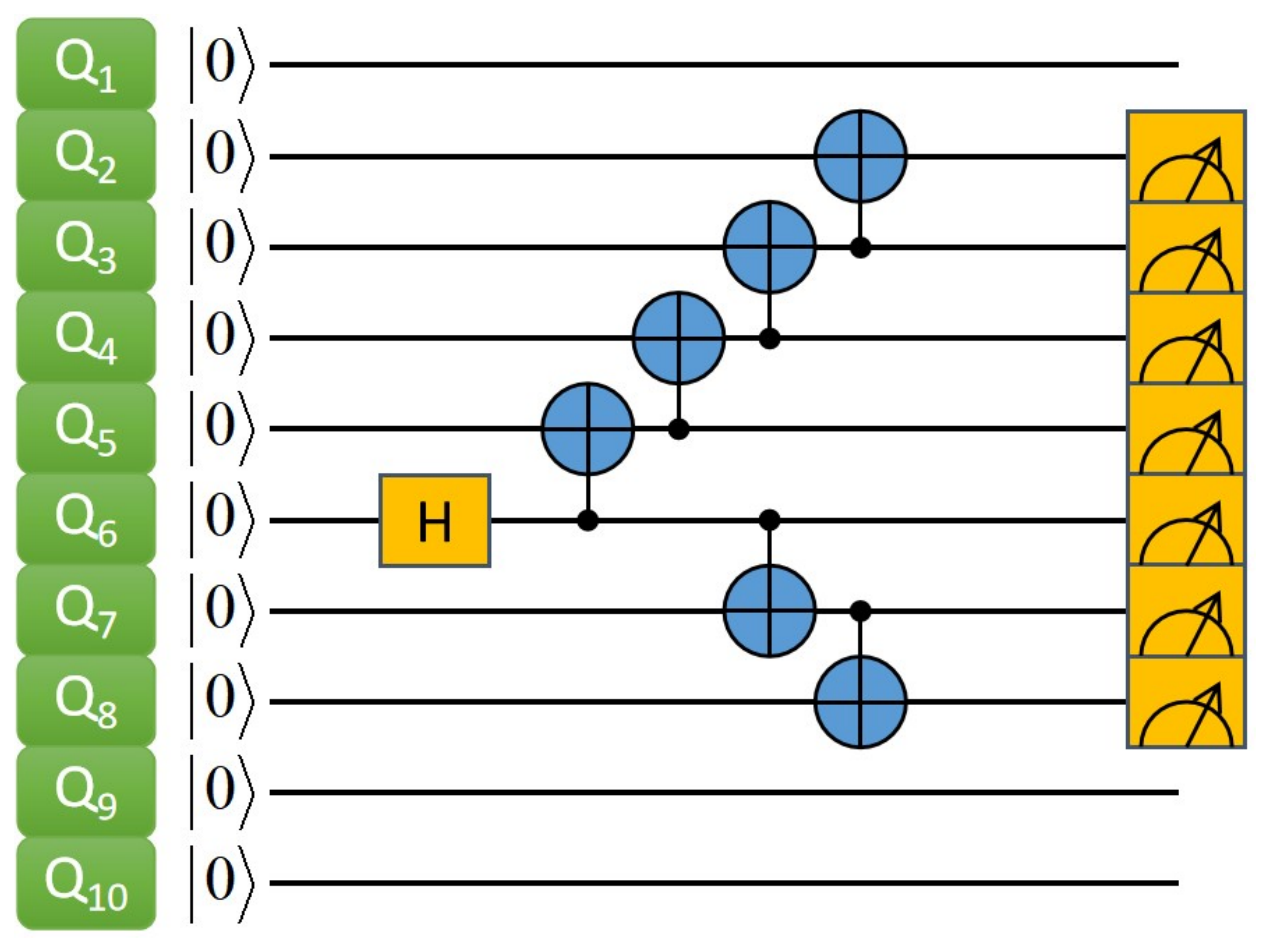}
\\
\vskip 0.5cm
\includegraphics[width=0.31\textwidth]{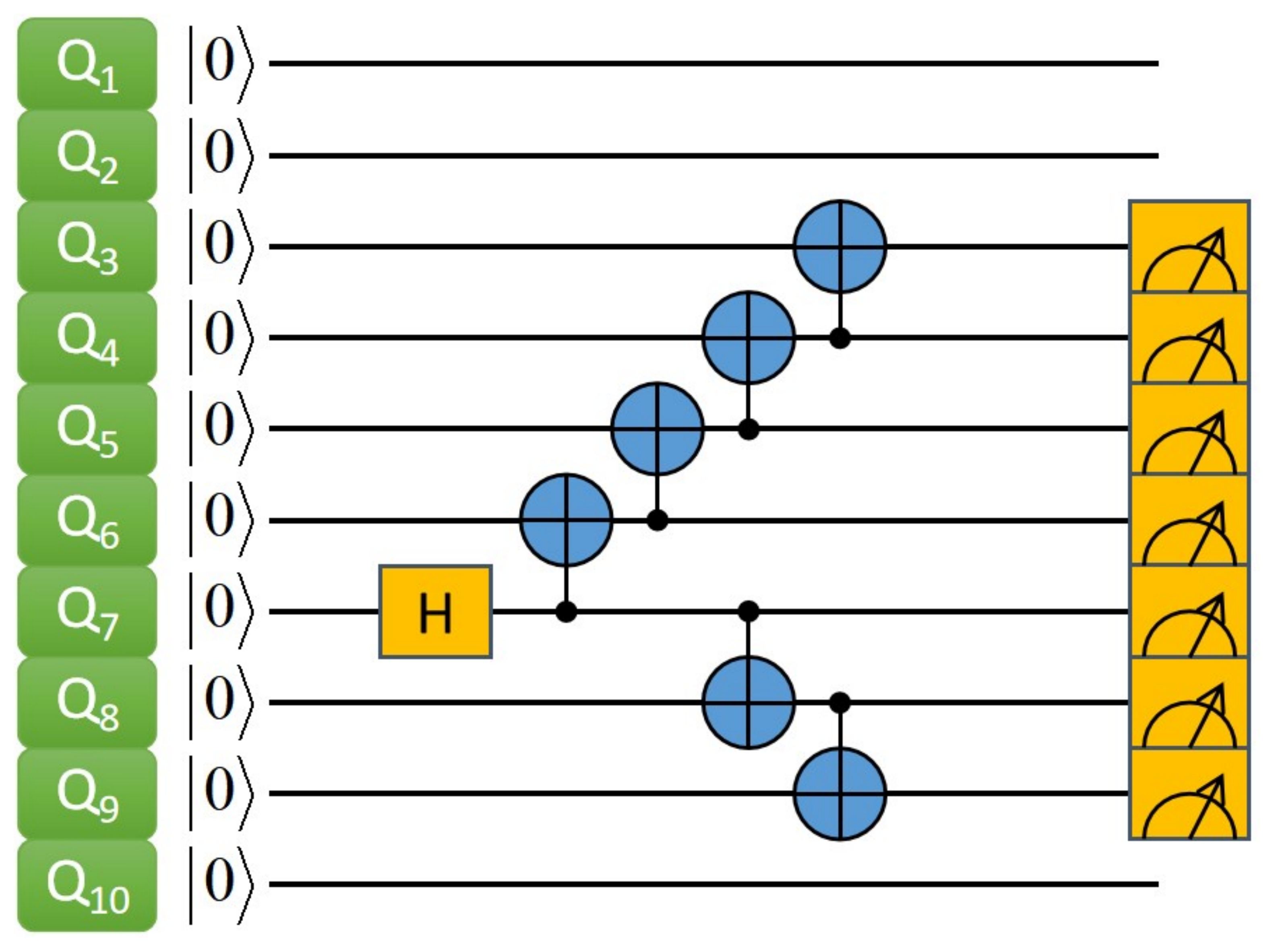}
~~~
\includegraphics[width=0.31\textwidth]{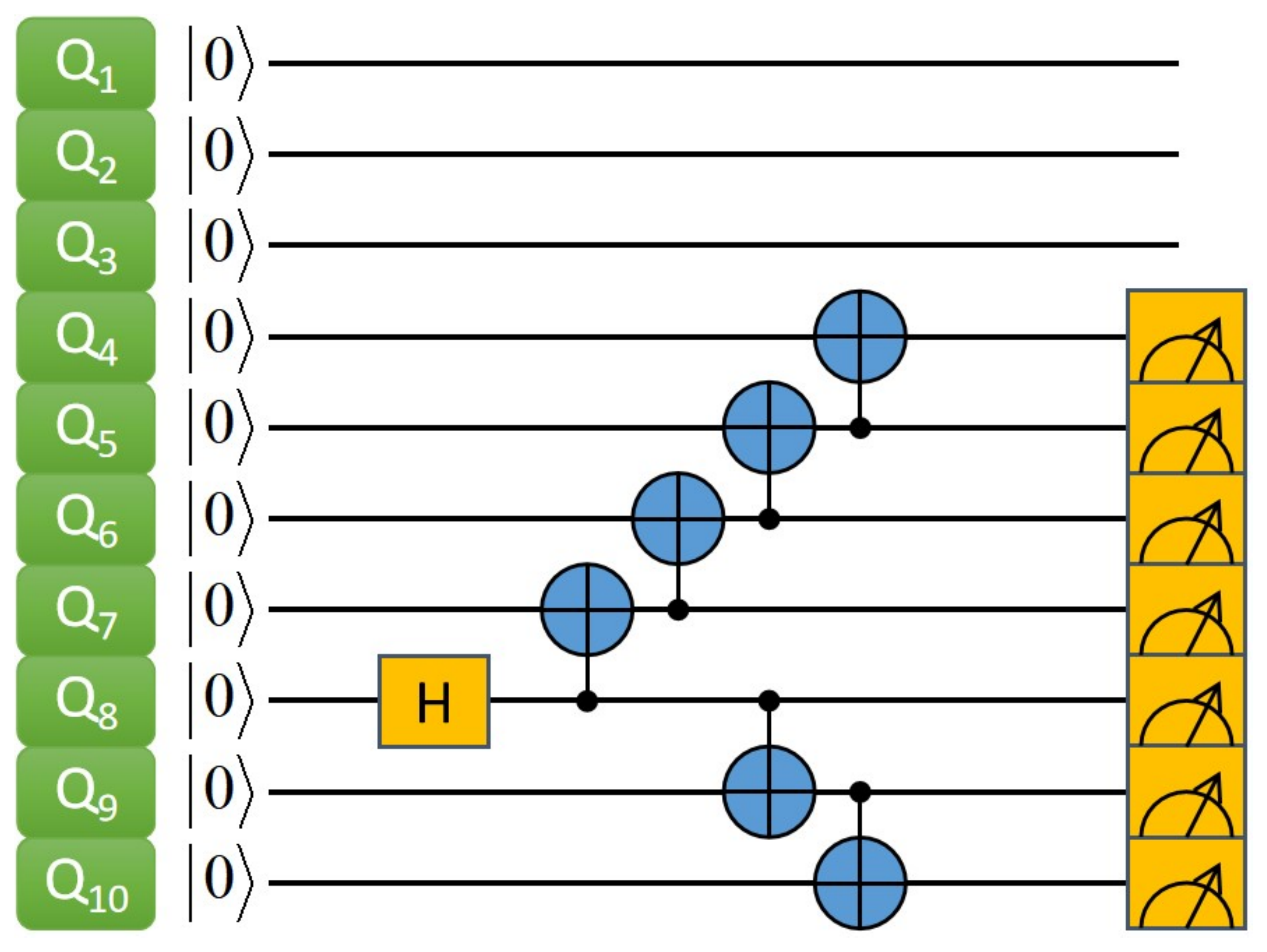}
\caption{Quantum circuits of 7-qubit GHZ state generation}
\label{7qubitGHZ}
\end{figure}

\begin{figure}[ht]
\centering
\includegraphics[width=0.31\textwidth]{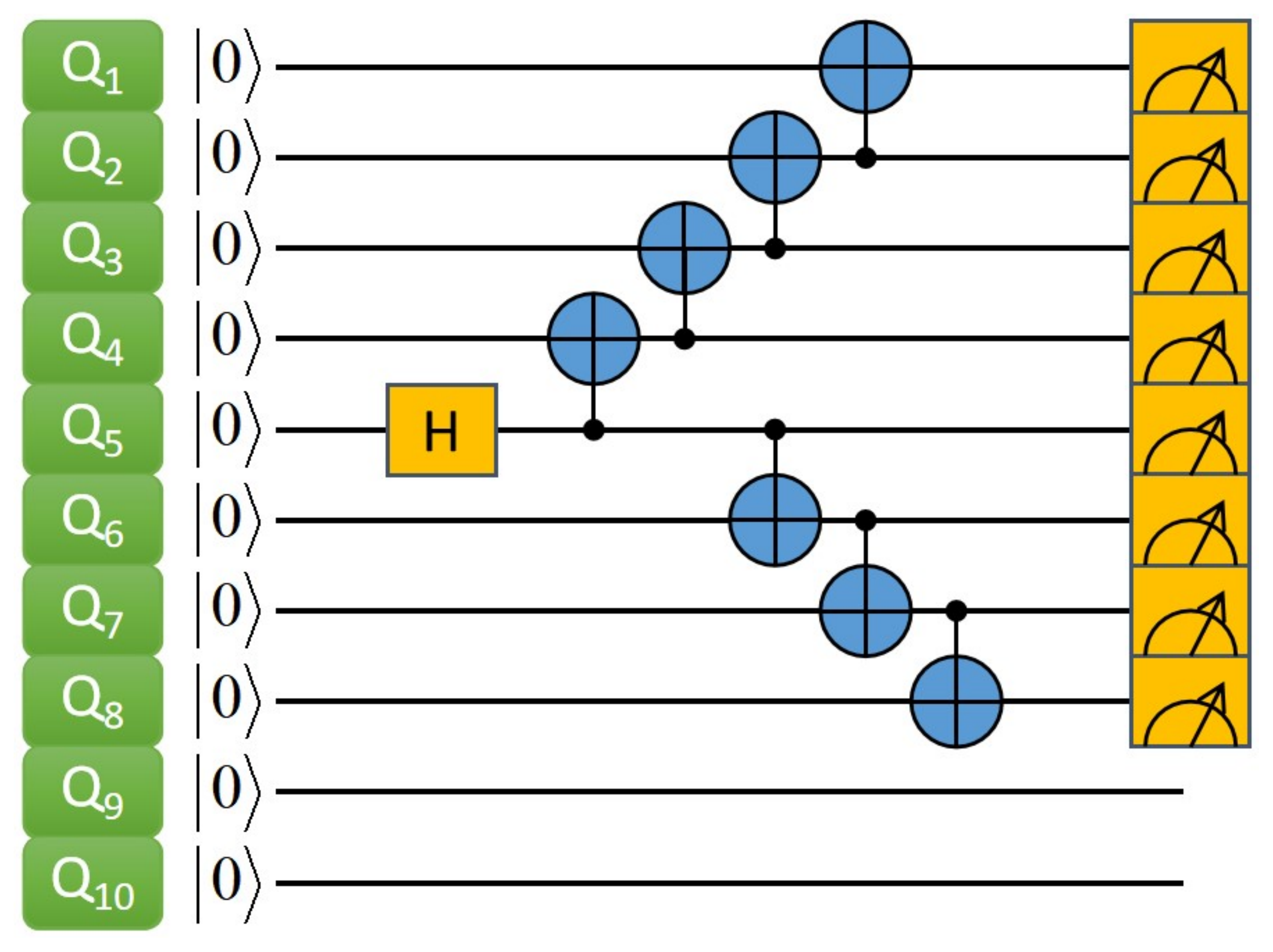}
~~
\includegraphics[width=0.31\textwidth]{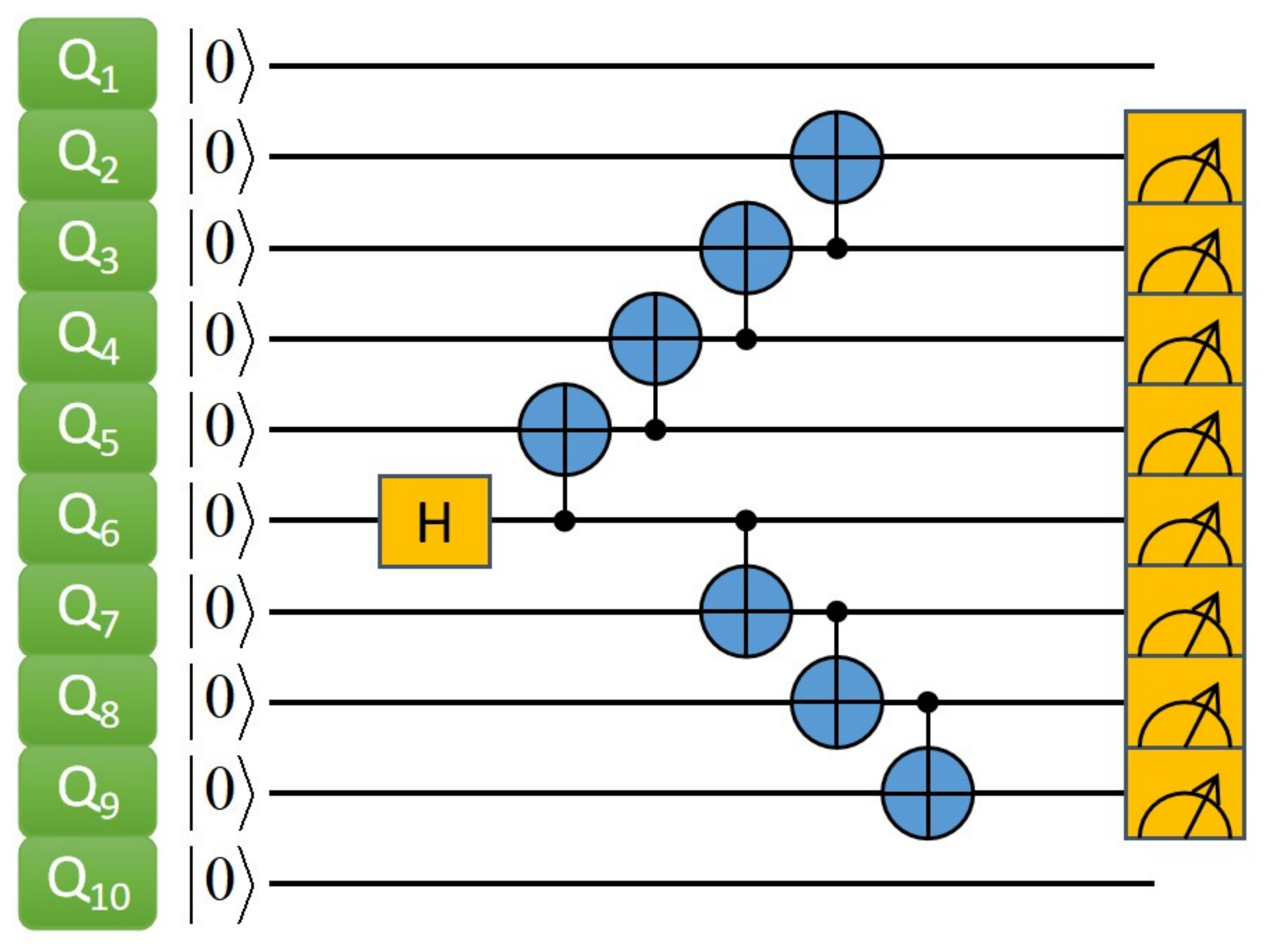}
~~
\includegraphics[width=0.31\textwidth]{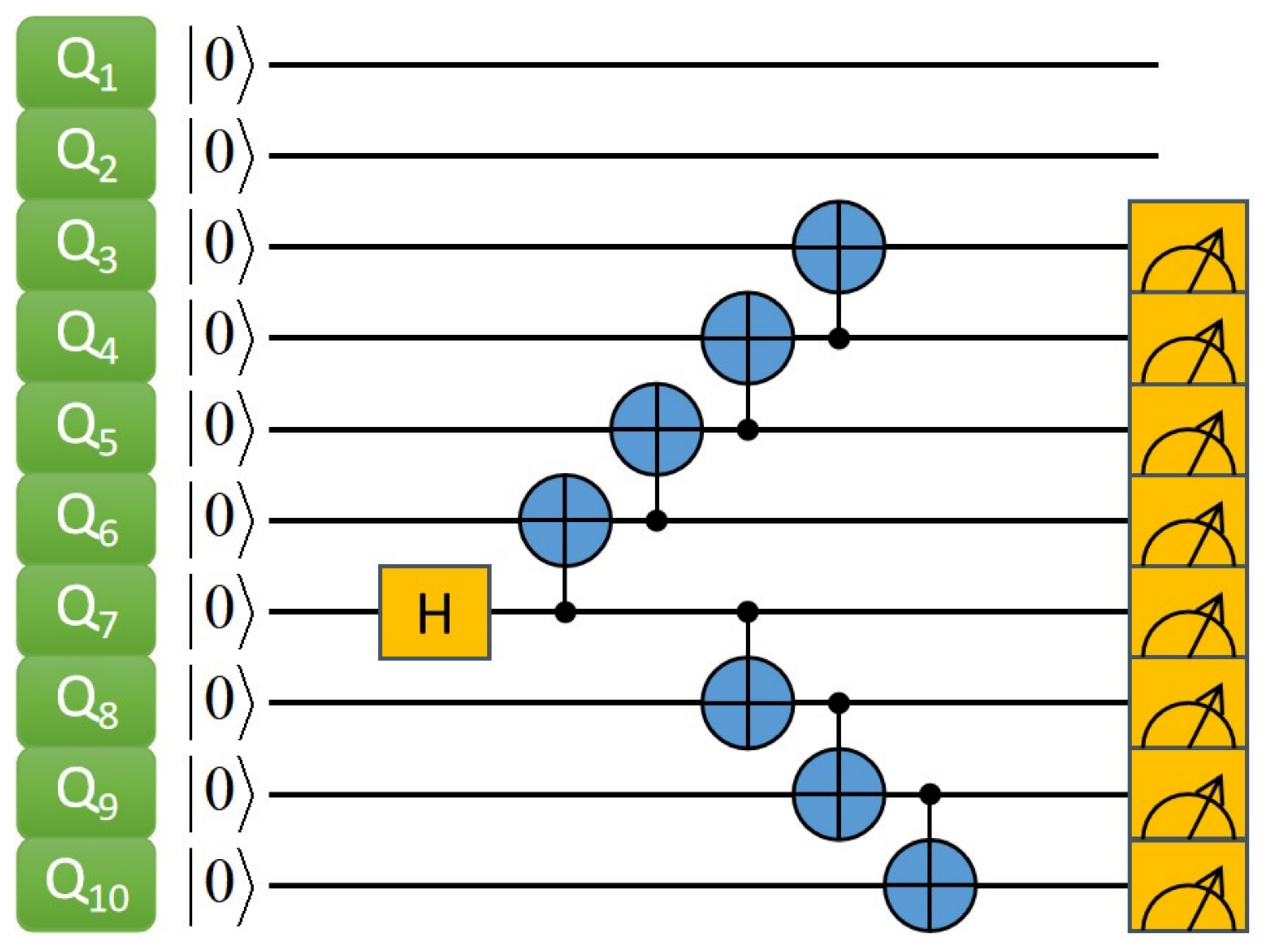}
\caption{Quantum circuits of 8-qubit GHZ state generation}
\label{8qubitGHZ}
\end{figure}

\begin{figure}[ht]
\centering
\includegraphics[width=0.31\textwidth]{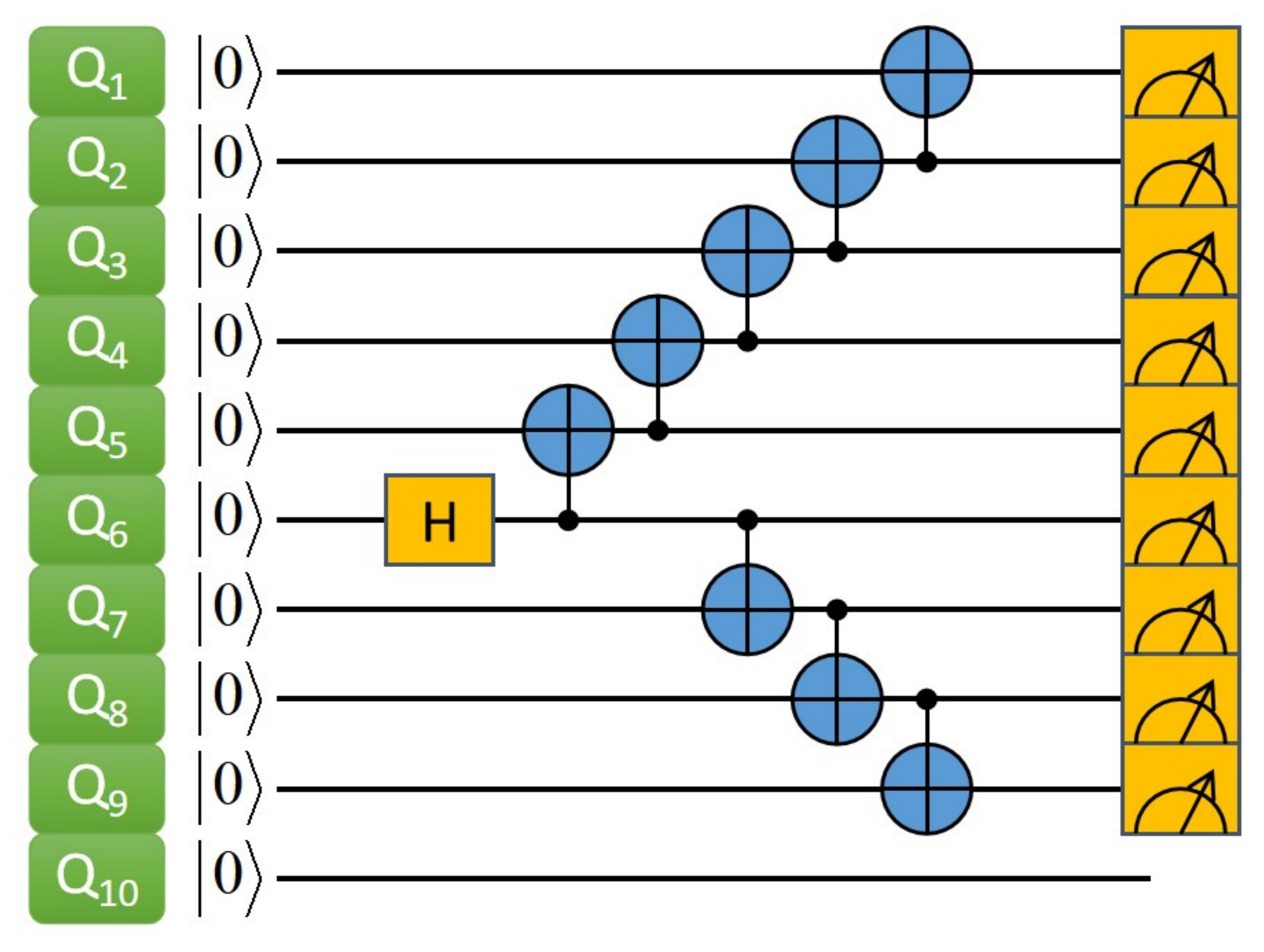}
~~
\includegraphics[width=0.31\textwidth]{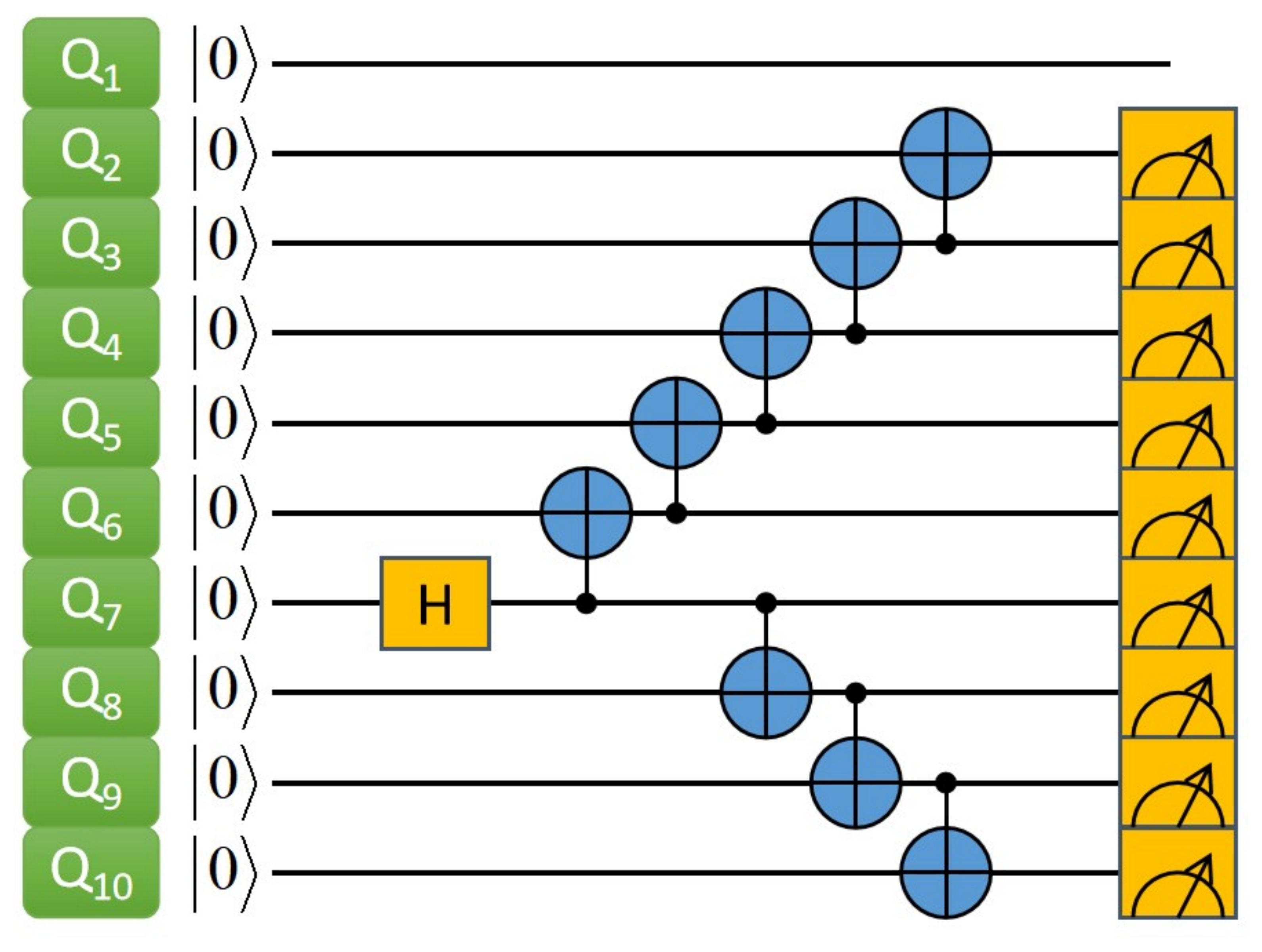}
\caption{Quantum circuits of 9-qubit GHZ state generation}
\label{9qubitGHZ}
\end{figure}

In this section, we show the quantum circuits of GHZ state generation with qubit number from 6 to 9, which are mentioned in the main text. See Fig.~\ref{6qubitGHZ}, Fig.~\ref{7qubitGHZ}, Fig.~\ref{8qubitGHZ}, Fig.~\ref{9qubitGHZ} for GHZ states of 6$-$9 qubits, respectively. Note that to avoid crosstalk, parallel quantum operations which involve two-qubit gates cannot be excuted for nearest-neighboring qubits. In the text, we use histogram to intuitively display the fidelity of different GHZ States, and the corresponding detailed fidelity data are shown in Tab.~\ref{GHZ_fid}
\begin{table}[ht]
    \centering
    \setlength{\tabcolsep}{2mm}{
    \begin{tabular}{|c|ccccc|cccc|} 
        \hline  $N$ & \multicolumn{5}{c|}{$N$=6} & \multicolumn{4}{c|}{$N$=7}\\
        \hline Constituent qubits  & $Q_{1}$-$Q_{6}$ & $Q_{2}$-$Q_{7}$ & $Q_{3}$-$Q_{8}$ & $Q_{4}$-$Q_{9}$ & $Q_{5}$-$Q_{10}$ & $Q_{1}$-$Q_{7}$ & $Q_{2}$-$Q_{8}$ & $Q_{3}$-$Q_{9}$ & $Q_{4}$-$Q_{10}$ \\
        \hline Population (\%) &93.8(6) &91.0(2) &89.0(3) &91.5(4) &89.0(6) & 89.3(2) & 95.5(3) & 89.3(5) & 92.4(4)\\
        \hline Coherence (\%) &88.4(2) &88.8(3) &89.1(2) &91.4(5) &77.7(4) & 88.8(4) & 91.0(3) & 88.3(4) & 81.1(5)\\
        \hline Fidelity (\%) &91.1(4) &89.9(2) &89.1(2) &91.4(2) &83.4(3) & 89.0(3) & 93.3(2) & 88.8(4) & 86.7(5)\\
        \hline
        
    \end{tabular}}
    \setlength{\tabcolsep}{4mm}{
    \begin{tabular}{|c|ccc|cc|c|} 
    \hline  $N$ & \multicolumn{3}{c|}{$N$=8} & \multicolumn{2}{c|}{$N$=9}& \multicolumn{1}{c|}{$N$=10}\\
        \hline Constituent qubits  & $Q_{1}$-$Q_{8}$ & $Q_{2}$-$Q_{9}$ & $Q_{3}$-$Q_{10}$ & $Q_{1}$-$Q_{9}$ & $Q_{2}$-$Q_{10}$ & $Q_{1}$-$Q_{10}$ \\
        \hline Population (\%) &86.8(2) &90.4(2) & 88.3(3) & 88.9(1) & 86.9(3) & 80.1(4) \\
        \hline Coherence (\%) &79.1(2) &73.6(1) & 73.5(4) & 75.2(3) & 76.9(5) & 75.6(2) \\
        \hline Fidelity (\%)  &82.9(1) &82.0(1) & 80.9(3) & 82.0(2) & 81.9(3) & 77.9(2) \\
        \hline
    \end{tabular}}
    \caption{Populations, coherences and fidelities of all GHZ states. On this processor, we have prepared all possible GHZ states with no less than 6 constituent qubits, and the corresponding fidelities are listed. The populations(coherences) are obtained by the method on the left(right)-hand side in Fig. 1(b) in the text. Each fidelity is obtained by averaging corresponding population and coherence.}
    \label{GHZ_fid}
\end{table}

\end{document}